\documentclass[12pt,preprint]{aastex}
\def \lp{\>\> .}

\def \arcsec{\hbox{$^{\prime\prime}$}}
\def \arcmin{\hbox{$^{\prime}$}}
\def \deg{$^\circ$}
\def \hr{$^h$}
\def \mn{$^m$}
\def \sc{$^s$}

\def \c2{cm$^{-2}$}
\def \cc{cm$^{-3}$}
\def \kms{km s$^{-1}$}

\def \kks{K kms$^{-1}$}

\def \av{A$_{\rm V}$}
\def \n2h{N$_2$H$^+$}

\def \nh2{n_{H_2}}
\def \nh1{n_{HI}}

\def \tw{$^{12}$CO}
\def \th{$^{13}$CO}
\def \ce{C$^{18}$O}
\def \h2{H$_2$}
\def \nh3{NH$_3$}

\def \Ls{$L_{\odot}$}
\def \Ms{$M_{\odot}$}

\def \be{\begin{equation}}
\def \ee{\end{equation}}
\def \bf{\begin{figure}}
\def \ef{\end{figure}}

\def \lp{\>\> .}

\begin{document}

\shorttitle{Molecular Gas in Taurus}
\shortauthors{Goldsmith et al.}

\title{Large--Scale Structure of the Molecular Gas in Taurus Revealed by High Linear Dynamic Range 
Spectral Line Mapping}

\author{Paul F. Goldsmith\altaffilmark{1}, Mark Heyer\altaffilmark{2}, Gopal Narayanan\altaffilmark{2}, Ronald Snell\altaffilmark{2}, Di Li\altaffilmark{3}, and Chris Brunt\altaffilmark{4}}
\altaffiltext{1}{Jet Propulsion Laboratory, California Institute of Technology, 4800 Oak Grove Drive, Pasadena CA, Paul.F.Goldsmith@jpl.nasa.gov}
\altaffiltext{2}{FCRAO, Department of Astronomy, University of Massachusetts, Amherst MA}
\altaffiltext{3}{Jet Propulsion Laboratory, California Institute of Technology, 4800 Oak Grove Drive, Pasadena CA}
\altaffiltext{4}{University of Exeter}

\begin{abstract}

We report the results of a 100 square degree survey of the Taurus
Molecular Cloud region in the J = 1 $\rightarrow$ 0 transition of
\tw\ and of \th.  The image of the cloud in each velocity channel includes $\simeq$
3$\times$10$^6$ Nyquist-—sampled pixels, sampled on a 20\arcsec\
grid.  The high sensitivity and large linear dynamic range of the maps
in both isotopologues reveal a very complex, highly
structured cloud morphology.  
There are large scale correlated structures evident in
\th\ emission having very fine dimensions, including filaments, cavities, and
rings.  The \tw\ emission shows a quite different structure, with particularly
complex interfaces between regions of greater and smaller column
density defining the boundaries of the largest--scale cloud structures.
The axes of the striations seen in the \tw\ emission from
relatively diffuse gas are aligned with the direction of the magnetic field.
We have developed a statistical method for analyzing
the pixels in which \tw\ but not \th\ is detected, which allows us to
determine the CO column in the diffuse portion of the cloud as well as
in the denser regions in which we detect both isotopologues.  
Using a column density--dependent model for the CO fractional abundance, we derive the mass
of the region mapped to be 2.4$\times$10$^4$ \Ms.  This is more than a factor of two
greater than would be obtained using a canonical fixed fractional abundance of \th\, 
and a factor three greater than would be obtained using this fractional abundance 
restricted to the high column density regions.  We determine that half the mass
of the cloud is in regions having column density below
2.1$\times$10$^{21}$ \c2.  The distribution
of young stars in the region covered is highly nonuniform, with the
probability of finding a star in a pixel with a specified column
density rising sharply for $N(H_2)$ = 6$\times$10$^{21}$ \c2.  We
determine a relatively low star formation efficiency (mass of young
stars/mass of molecular gas), between 0.3 and 1.2 percent, and an average
star formation rate during the past 3 Myr of 8$\times$10$^{-5}$ stars yr$^{-1}$.

\end{abstract}

\setcounter{footnote}{0}

\section{INTRODUCTION}
\label{introduction}

The close association of young stars and concentrations within
molecular clouds indicates that stars form in cloud cores, which are
regions of increased density within the bulk of molecular clouds
\citep[cf.][]{beichman1986}.  While the evolution from cloud core to
protostar is dominated by gravity, the physics controlling the process
in which the cores themselves, and the clouds in which they are
embedded, are formed and evolve is still quite controversial.  While on
the scale of pc to tens of pc molecular clouds are close to
satisfying virial equilibrium between gravitational and kinetic
energies, the significance of this equality is not entirely clear.
Furthermore, the role of magnetic field, while often postulated to be
significant, remains uncertain \citep{shu1987, heiles2005}.  
Finally, the formation of molecular clouds themselves, and their lifetime, 
remains very much a matter of discussion \citep[e.g.][]{hartmann2001}

Molecular clouds may be formed by compression of atomic gas, with the
increased density and extinction enhancing the formation rate of
molecules, starting with \h2, for which self--shielding enables the
buildup of a substantial fraction of the total hydrogen density even
when the visual extinction \av\ is only a fraction of a magnitude.  It
has also been suggested that the large molecular cloud presence in
galactic spiral arms is the result of the agglomeration of molecular
material existing in the interarm region, as discussed by
\cite{pringle2001}.  While one viewpoint has held that molecular
clouds have relatively long lifetimes, and are disrupted only by the
energy injected by massive star formation and evolution,
another picture is that molecular clouds are relatively transient
objects, with the denser regions representing only turbulent
fluctuations of density rather than well-defined gravitationally bound
condensations \citep[see e.g. review by][]{vazquez2007}.

These issues have been discussed on global scale, addressing the
distribution of clouds and the apportioning of molecular and atomic
gas in the Galaxy. They are also very relevant to studies of specific molecular
cloud complexes, with one of the best--studied of these being that in
Taurus.  The structure of the interstellar gas in atomic and molecular
form, the stellar population, the issue of star formation rate, and
the role of different physical processes have all been the subject of
numerous papers focused on the Taurus region, primarily because its
proximity \citep[140 pc;][]{elias1978}\footnote{This value, from
\citet{elias1978}, is so entrenched in the literature that we will use it
despite the plausible suggestion by \cite{hartigan2003} that the
distance should be reduced by about 10\%, to $\sim$126 pc.} allows
very detailed studies of the morphology of the gas and the
relationship between gas and stars.  The sheer volume of the data that
have been obtained and the number of analyses that have been carried
out preclude giving a complete listing of the references to Taurus, so
we will have to be selective rather than comprehensive, recognizing
that we may have omitted many valuable contributions.

The very closeness of Taurus means that available instrumentation,
particularly at radio frequencies, has faced a challenge to cover the
entire region with angular resolution sufficient to reveal the
morphology of the gas.  The result has been that previous large--scale
surveys of molecular line emission at millimeter wavelengths have been
limited to quite low angular resolution \citep{ungerechts1987}.  
The survey of \cite{ungerechts1987} covers essentially all of 
Taurus and part of Perseus, but
the 30\arcmin\ angular resolution of the map (obtained by averaging
multiple telescope pointings to obtain a larger effective beam size) yields only
3000 pixels in the 750 square degree region mapped.  The pixel size
corresponds to a linear size of 1.2 pc at a distance of 140 pc, which
is sufficiently large to blur out structure at important astrophysical
scales.  In fact, the maps of \cite{ungerechts1987}, while delineating
the large--scale structure quite well, show an almost complete absence
of fine detail.  This is in part due to the use exclusively of \tw,
which is sufficiently optically thick that significant variations in
column densities can be entirely hidden, as well as to the low angular
resolution.

There have been a number of investigations of molecular gas in the
Taurus region with higher angular resolution, but these have typically
been limited to small subregions within the overall gas distribution.
These studies, with $\simeq$1\arcmin\ to 2\arcmin\ angular resolution
include a few thousand to $\simeq$ 30,000 spatial pixels 
\citep{schloerb1984,duvert1986, heyer1987,mizuno1995}.  
These studies, with the  combination of higher angular resolution and use of the J =
1$\rightarrow$0 transition of \th\, do reveal considerable structure
in the molecular gas, but have not elucidated its relationship to
larger--scale features in the molecular gas distribution.

A number of other studies have utilized yet higher angular resolution
and different tracers to probe gas having different characteristic
properties over limited regions.  Some examples include
\cite{langer1995} employing CCS, \cite{onishi1996} and
\cite{onishi1998} using \ce, \cite{onishi2002} using H$^{13}$CO$^+$,
and \cite{tatematsu2004} employing N$_2$H$^+$.  Many individual cores
have been observed in ammonia, a tracer in which they appear relatively
well--defined, as indicated by compilation of \citet{jijina1999}.
Most of the regions covered by these studies have been pre-selected
based on the large--scale surveys discussed above.  In these maps, we
see indications of finer--scale structure, but the emission is
generally quite spatially restricted compared to that seen in the more
abundant isotopologues of carbon monoxide.

In this paper we present the initial results from a large--scale high
angular resolution study of the Taurus molecular clouds using \tw\ and
\th.  The data cover approximately 100 square degrees on the sky
(11.5\arcdeg\ in R.A. by 8.5\arcdeg\ in decl.) corresponding to a region 28 pc
by 21 pc.  The reduced maps include 3.2$\times$10$^6$ Nyquist--sampled
pixels in each isotopologue, with pixel size 20\arcsec\, corresponding
to 0.014 pc.  The linear dynamic range (LDR, defined as map size
divided by Nyquist--sampled interval) of the maps thus exceeds 1000, which is the
largest of any molecular cloud study carried out to date.  The good
angular resolution and large LDR together allow us to examine in
detail the relationship between the relatively fine structures seen,
especially in \th, with the large--scale distribution of the molecular
material, the young stars in the region, and the magnetic field.

The region of Taurus studied here has been observed using a variety of
other tracers.  The Leiden/Dwingeloo 21 cm study \citep{burton1994}
traced the atomic hydrogen in this direction, but with an angular
resolution of 35\arcmin.  One investigation \citep{shuter1987} used
the Arecibo radio telescope having an angular resolution of 4\arcmin,
but included only $\sim$ 1300 positions to probe the self--absorption
seen in the 21 cm HI line.  This cold atomic hydrogen appears to be
associated with molecular gas \citep{li2003, goldsmith2005}, but the
limited sampling of Shuter et al.\ does not reveal much about its morphology.  
The far--infrared emission from Taurus has been studied by
\cite{abergel1995}, who also compared it to moderate resolution maps
of \tw\ J = 1$\rightarrow$0 emission.  The dust column density
distribution has been examined by \cite{padoan2002} and does bear a
quite close resemblance to the integrated intensity of \th\, and thus
to the column density of gas in relatively high extinction regions.

We discuss the observations and data reduction procedure in \S \ref{observations}.  
Derivation of the column density in the different
portions of the maps is presented in \S \ref{coldens}, in which we
also discuss the distribution of column density and mass in the
region.  We present a brief discussion of the large--scale gas kinematics in
\S \ref{kinematics}.  
We address the relationship of the molecular material and the magnetic field in
\S \ref{magnetic}, and discuss the relationship of the gas and
the young stars in the region in \S \ref{youngstars}.  
We discuss some of the interesting features of
the morphology of the gas in \S \ref{morphology}.  
We summarize our results in \S \ref{summary}.


\section{OBSERVATIONS}
\label{observations}

The observations were taken between 2003 November and 2005 May using
the 13.7m radome--enclosed Quabbin millimeter wave telescope.  The 32 pixel
SEQUOIA focal plane array\footnote{A 16 pixel single--polarization
version of the array is described in \cite{erickson1999}.  }  receiver
observed the J = 1$\rightarrow$0 transition of \tw\ and \th\
simultaneously.  Since the receiver uses amplifiers for the first
stage, there is no issue of the sideband gain uncertainty and its
effect on calibration.  Sixteen pixels are arranged in a 4 x 4 array
in two orthogonal linear polarizations.  The main beam of the antenna
pattern had a full width to half maximum angular width of 45\arcsec\
for \tw\ and 47\arcsec\ for \th.

The data were obtained using an on--the--fly (OTF) mapping technique.
A standard position was observed using position switching several times 
per observing session to verify calibration consistency.
Details of the data--taking, data reduction, and calibration procedures 
are given by \cite{narayanan2007}.
The signals from a band of frequencies around each spectral line were
sent to an autocorrelation spectrometer with 1024 lags covering 25 MHz
for each spectral line.  The lag spacing of the spectrometer system
corresponds to 0.068 \kms\ for \th\ and 0.065 \kms\ for \tw.  
The data cube of each isotopologue employed in the subsequent analysis included 76 spectral
channels for \th\ and 80 channels for \tw\ covering approximately -5 \kms\ to +14.9 \kms\, and thus
included 2.4$\times$10$^8$ voxels. 

As discussed in detail by \cite{narayanan2007}, the overall quality of the data was
excellent.  
After calibration and combination of the 30\arcmin\ by 30\arcmin\ submaps which were the
units in which the data was taken, the data were
resampled onto a uniform grid of 20\arcsec\ spacing, which is very
close to the Nyquist sampling interval $\lambda/2D$ for the 13.7 m
diameter telescope operating at a wavelength of 2.6 mm.  The images
produced by the combination of the submaps and regridding were 2069
pixels in RA by 1529 pixels in decl., thus comprising 3,163,501
spatial pixels resampled onto a uniform 20\arcsec\ grid.
The final data set has a well--behaved
distribution of noise with a mean rms antenna temperature equal to
0.125 K for \th\ and 0.28 K for \tw\ in channel widths of 0.27 \kms\
and 0.26 \kms, respectively.

%
\begin{figure}[!htbp]
\includegraphics[scale=0.85]{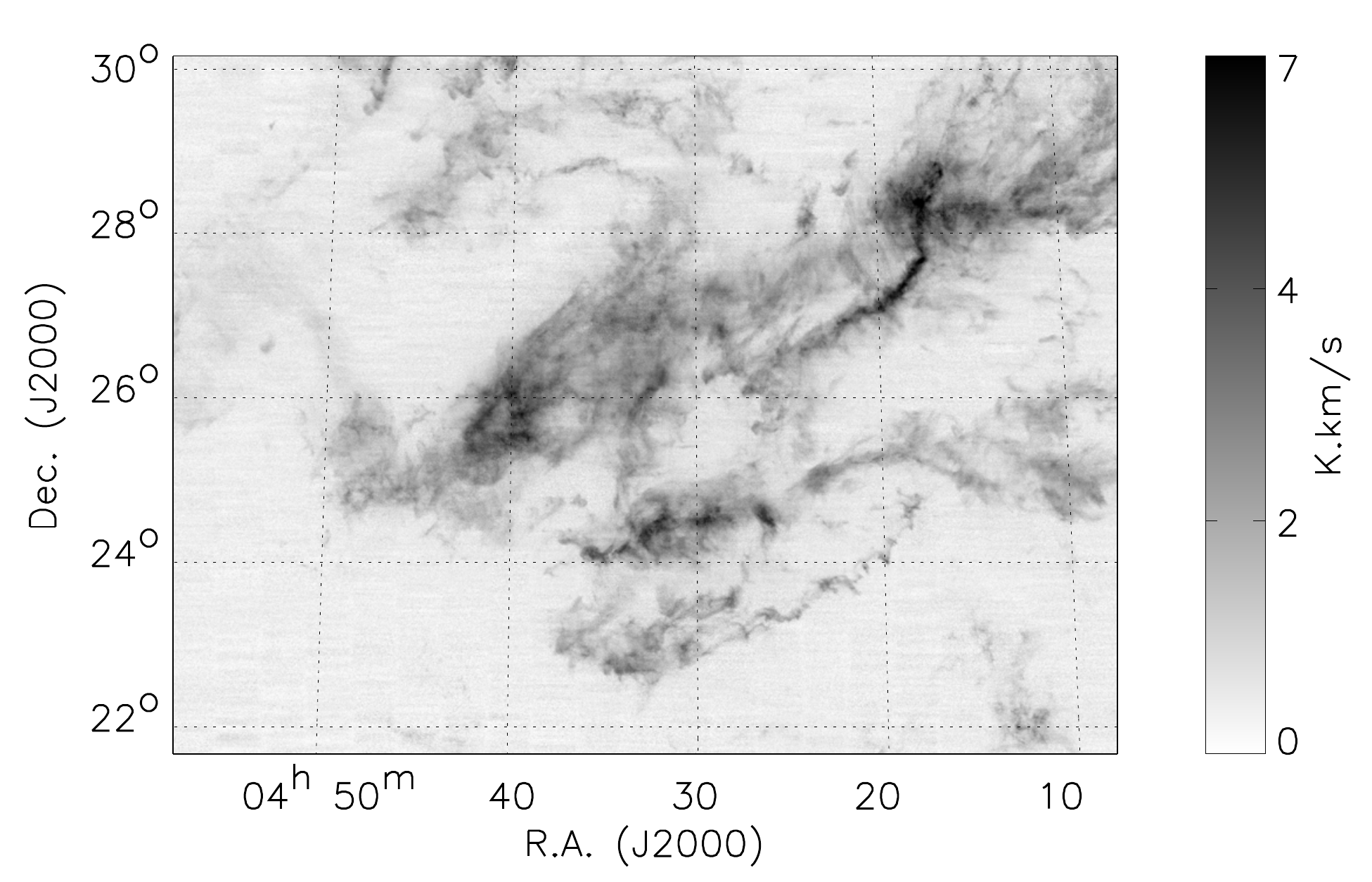}
\caption{\label{13co_tint} Antenna temperature of the \th\ J =
1$\rightarrow$0 transition integrated over the velocity range 2 \kms\
to 9 \kms.  The scale is shown in the bar at the right; values have not been
corrected for antenna efficiency.}
\end{figure}

We show the basic \th\ data in Figure \ref{13co_tint}, which gives the
intensity of the \th\ J = 1$\rightarrow$0 transition integrated over
the velocity range 2 \kms\ to 9 \kms.  This interval encompasses
almost all of the emission in the Taurus region, with the exception of
some isolated areas with gas at $\simeq$ 10 \kms, which may well not
be associated with Taurus, and a limited amount of emission in the 
velocity range 1 \kms\ to 2 \kms.  
Figure \ref{12co_tmax} displays the \tw\ J
= 1$\rightarrow$0 peak emission within this same velocity interval.
Note that in both of these figures, the emission is not corrected for
the antenna efficiency.  
\cite{narayanan2007} present images of the emission of both isotopologues 
in 1 \kms\ bins covering the range 0 \kms\ to 13 \kms.

It is evident that the \tw\ is detectable over a significantly larger
area than is the \th.  Particularly in the northeast portion of the map,
we see very extended \tw\ emission, where there is relatively little \th. 
There are also two interesting regions of quite strong \tw\ emission,
at 4\hr22\mn +28\deg30\arcmin\ and 4\hr48\mn +29\deg40\arcmin, which are
among the warmest regions observed, and yet which do not show up as significant
local maxima in the \th\ (and hence column density).  In general, the
warmer gas as traced by \tw\ is seen in regions of high column density, but
the amount of structure seen in the optically thick \tw\ with our angular
resolution, sampling, and sensitivity, is very impressive.  

%
\bf[!htbp]
\begin{center}
\includegraphics[scale=0.85]{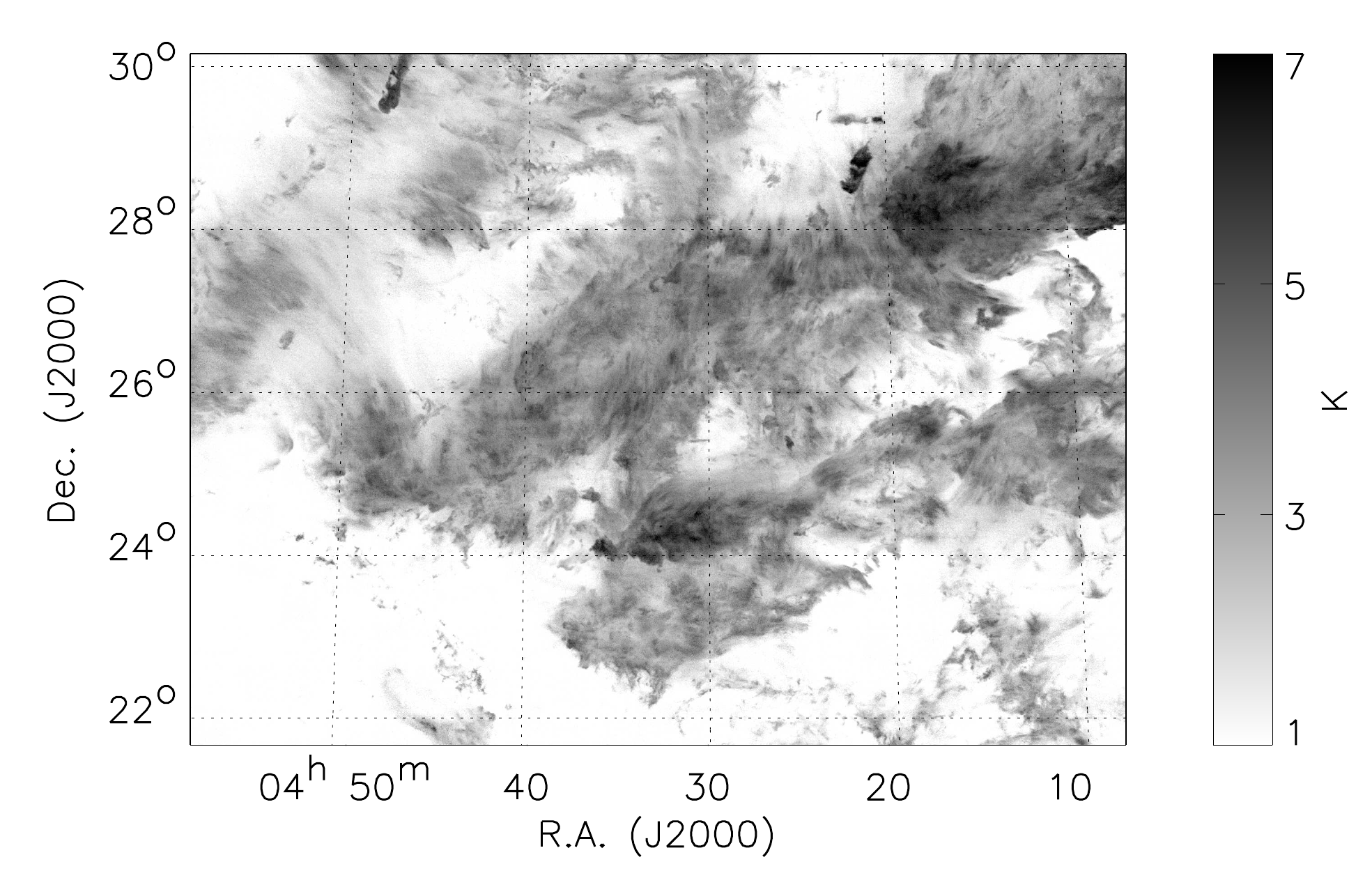}
\caption{\label{12co_tmax} Maximum antenna temperature of the \tw\ J =
1$\rightarrow$0 transition over the velocity range 2 \kms\ to 9 \kms.
The antenna temperature has not been corrected for the antenna
efficiency.}
\end{center}
\ef

\section{COLUMN DENSITY, COLUMN DENSITY DISTRIBUTIONS, AND CLOUD MASS}
\label{coldens}

\subsection{Mask Regions}

In order to facilitate analysis of the data to determine column
densities, we have broken the Taurus data up into 4 regions, according
to the detection or nondetection of \tw\ and \th.  The detection
thresholds are defined by the requirement that the integrated
intensity over the velocity range extending from 0 \kms\ to 12 \kms\
be a minimum of 3.5 times larger than the rms noise in an individual
pixel over this 12 \kms\ velocity interval.  The median values are
$\sigma_{T_{int}}$ = 0.18 \kks\ for \th\ and $\sigma_{T_{int}}$ = 0.40
\kks\ for \tw.  Since the peak values of the integrated intensity are
6 \kks\ for \th\ and 18 \kks\ for \tw, the peak integrated intensities
are 30 to 50 $\sigma_{T_{int}}$.

We define mask 0 to be the region in which neither \tw\ nor \th\ is
detected, mask 1 to be the region in which \tw\ is detected but \th\
is not, mask 2 to be the region in which both isotopologues are
detected, and mask 3 to be the region in which \th\ is detected but
\tw\ is not.  The different regions and the number of pixels in each
are given in Table \ref{maskregiontable}.


\begin{deluxetable}{ccc}
\tablewidth{0pt}
\tablecaption{\label{maskregiontable} Mask Regions in the Taurus Map}
\tablehead{\colhead{Mask Region}	&\colhead{Characteristics}	&\colhead{Number of Pixels}\\} 
\startdata
0	&neither \tw\ nor \th\	& \phm{1,}944,802\\
1	&\tw\ but not \th\	& 1,212,271\\
2	&both \tw\ and \th\	& 1,002,955\\
3	&\th\ but not \tw\	& \phm{1,00}3,473\\
\enddata
\end{deluxetable}

The average spectra of mask 0, mask 1, and mask 2 regions are shown in
Figure \ref{3masks_spectra}.  These profiles are valuable for deducing
general characteristics of the regions, but it must be kept in mind
that the characteristics of the average profile are quite different
from those of individual profiles.  The difference is primarily due to
systematic velocity shifts across the cloud; these result in the
average spectra being much weaker and broader than individual spectra.
The line width of the averaged mask 1 spectra is close to a factor of
2 greater than the average line width of spectra in this region.  For
mask 2, the ratio is $\simeq$1.5.  Along with this, the peak
intensities are much weaker than those seen in individual spectra or
even in spectra averaged over a restricted region.  Consequently, in
determining characteristics of the molecular gas, we have used
individual spectra wherever possible to derive physical quantities.

%
\bf[!htbp]
\includegraphics[scale=0.8]{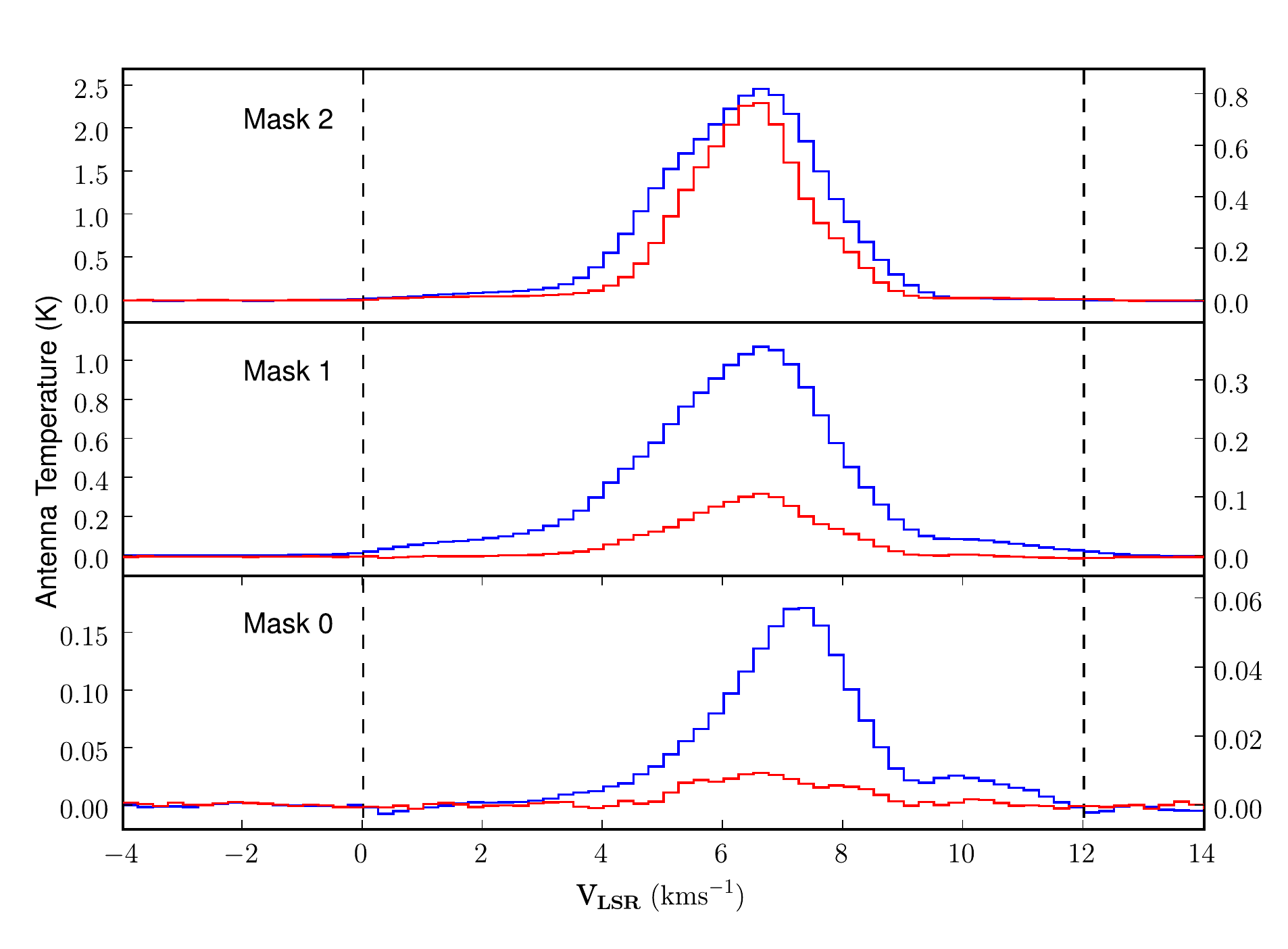}
\caption{\label{3masks_spectra} Spectra averaged in each of the three
significant mask regions of the Taurus map, arranged from mask 0
(bottom) to mask 2 (top).  In each panel, the \tw\ J= 1$\rightarrow$0
spectrum is the more intense (blue), while the weaker \th\ spectrum
is plotted in red.  The antenna temperature scale for the \tw\ is on the
left hand side, while that for the \th\ is on the right hand side, expanded 
by a factor of three relative to that of \th.
The dashed vertical lines delineate the velocity range used to define
the material in the Taurus region.
}  
\ef

As expected, the lines are strongest in mask 2.  The \tw\ to \th\
ratio at the line peak in mask 2 is just over 3, consistent with relatively high
optical depth in the more abundant isotopologue.  We do see that when
an average over $\sim$ 10$^6$ pixels of mask 1 is formed, we readily
see emission in \th\ as well as \tw.  The ratio of peak intensities is
significantly larger in mask 1 than in mask 2.  The value, about 10, is 
still much less than the presumed abundance ratio [\tw]/\th],  suggesting that the \tw\
in mask 1, while optically thick, typically has lower opacity than in
mask 2.

The mask 0 \tw\ and \th\ spectra show two or three 
peaks, including velocities for which the emission in mask
2 is very weak compared to that in the range of the peak emission, 5
\kms\ to 8 \kms.  In particular, the 10 \kms\ emission feature comes
from a fairly extended region in the northern portion of our map, but
is so weak that only when averaging over modest-sized ($\sim$1 square
degree) regions in mask 0 can it be detected. 
Emission in this velocity range can be quite clearly seen in the mask 1
spectrum, but hardly can be detected in mask 2.  This is consistent with
it being relatively low average column density material, which is extended over quite
large areas.
Thus, even in what we consider largely
``empty'' regions between the major, well--known subunits of the
Taurus molecular cloud complex, there is molecular gas.  This is discussed
further in the following section.  The overall composition of the mask
0 region, particularly the presence of atomic gas, is the subject of another study.

The mask 0, mask 1, and mask 2 regions have close to equal numbers of
pixels.  Their distribution, however, is very different.  Figure
\ref{mask_regions} shows the four mask regions.  It is evident that
the mask 1 predominantly surrounds mask 2, which is consistent with
the expectation that both isotopologues are detected in the regions of
highest column density (mask 2) while in the periphery of these
regions we detect in individual pixels the \tw\ but not the \th\ emission.

%
%
\bf[!htbp]
\includegraphics[scale=0.8]{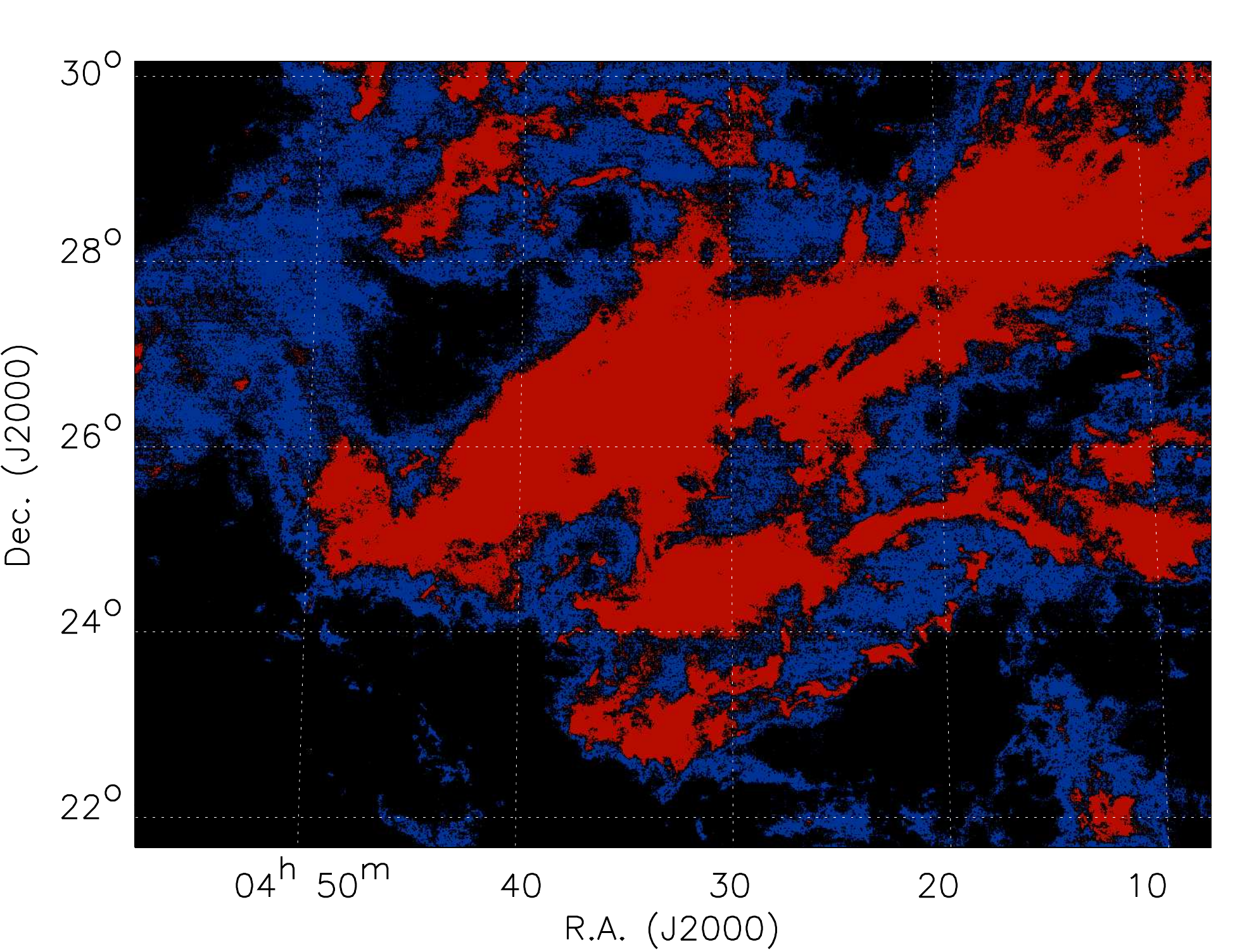}
\caption{\label{mask_regions} Image showing the mask regions (see text for definitions) in
the Taurus molecular cloud.  Mask 0 is shown in black, mask 1 in blue, and mask 2
in red.  The relatively few mask 3 pixels are not shown.}  
\ef

The pixels in mask 3 are unusual inasmuch as they exhibit detectable
\th\ emission but not \tw.  There are evidently very few such pixels
($\simeq$ 0.1\% of the total), although this number is considerably larger than
would be expected purely on the basis of Gaussian noise statistics.   
On close inspection of these spectra, it appears that
the problem is due to very low level baseline imperfections partially
canceling the \tw\ integrated intensity, resulting in a
``non--detection'' of this isotopologue.  We thus ignore the mask 3
pixels in further analysis of the emission from Taurus.

\subsection{Calculation of the Column Density}

We wish to exploit the large linear dynamic range of our map to
examine the structure in the column density, and thus wish to
determine the column density for as many pixels as possible.  This is also important
for accurately determining the total molecular mass of the region.  In
what follows we divide the problem into two parts.  The first is
determination of the carbon monoxide column density.  While subject to
its own uncertainties due to excitation, optical depth, and limited
signal to noise ratio, we can carry out this
step of the analysis based only on data in hand.  The second step is
conversion of the carbon monoxide column densities to molecular
hydrogen column densities, and finally to total cloud mass.  This is
evidently dependent on the processes which determine the fractional
abundance of the various isotopologues observed.  Since the additional
uncertainties in the second step are large, we present results
first in terms of the carbon monoxide distribution and subsequently
give results for the molecular hydrogen distribution and the total
molecular mass.  This second step should benefit significantly from
combination of our data with dust column density determined from
e.g.\ 2MASS data.  This effort is in progress and will be reported in a
subsequent publication.

\subsubsection{Carbon Monoxide Column Density}
\label{col_dens_calc}

The three different different regions of the cloud, defined by the detectability
of each isotopologue, require different schemes to determine the carbon 
monoxide column density.
We ignore mask 3 in determining the column density and to the mass of
the cloud as we cannot readily correct for the artificial
non--detections of \tw\ (discussed above).  Its extremely small area
and weak \th\ emission make its contribution negligible.

Mask 2 represents the portion of the cloud that is most conventional
in terms of column density determination.  Since we have both \tw\ and
\th\ in each pixel, we determine the kinetic temperature from the peak
value of the \tw\ (with appropriate correction for antenna
efficiency).  
Here (as well as for other mask regions), we use the maximum antenna temperature of \tw\ in the velocity interval between 0 \kms\ and 12 \kms.
The kinetic temperature is distributed from 3 K to 21 K,
but with the vast majority of positions having kinetic temperatures
between 6 K and 12 K.  Since mask 2 is the densest portion of the
cloud, we assume that the \th\ levels are populated in LTE at the
kinetic temperature, but we calculate a nominal value for the optical
depth from the ratio of the peak \th\ and \tw\ intensities using the
usual equation of radiative transfer in a uniform medium.  
We assume a \tw\ to \th\ abundance ratio of 65, very close to the average
value for local clouds found by \cite{langer1993}.
We use the value of optical depth obtained to make a saturation correction to the \th\ column
density derived assuming optically thin emission, with the usual
formula  
\be 
N(^{13}\rm{CO~corrected}) =
N(^{13}\rm{CO~assumed~optically~thin)}\frac{\tau}{1 - exp(-\tau)}\lp
\ee

Mask 1 presents the greatest challenge in terms of column density
determination since it encompasses approximately one third of the area mapped and
has reasonably strong \tw\ emission.  However, since the \th\ is not
detected in individual pixels, we need a different scheme to extract
the column density.  We have developed a statistical approach, which
should be applicable to other large maps in which only the more
abundant isotopologue is detected in individual pixels.  The procedure
assumes that the \tw\ is optically thick at its peak, and that the
value of the antenna temperature can directly be converted to the
excitation temperature of the \tw.  
Since mask 1 points lie at the periphery of the regions of high extinction
and greater molecular column density (as witnessed by the detection of \th\ in
each mask 2 pixel), they encompass lower column density gas which is 
presumably characterized by lower volume density.
Therefore we cannot assume that LTE applies as it
does in mask 2.  Approximately half of the mask 1 positions
have an excitation temperature $\le$ 7.5 K, and if in LTE the gas
would have to be unusually cold.  
It is thus reasonable to assume that this gas is subthermally excited.  
To analyze positions in mask 1 we use a simple
excitation/radiative transfer analysis employing a spherical cloud
large velocity gradient (LVG) code to compute the line intensities
\citep[e.g.][]{snell1981, goldsmith1983}.  We are using an LVG model
largely as a tool to characterize the effect of trapping, which is
important for excitation of CO at lower density.  We do not
believe it necessarily represents any statement about the detailed kinematics
of the gas.  The sensitivity of our results to the details of the velocity 
field should be quite small.

We have assumed that the kinetic temperature of the mask 1 region is
uniformly 15 K, somewhat higher than well-shielded dense gas, which is
plausible in view of increased heating in the peripheral regions surrounding
regions of high extinction. \citep[e.g.][]{li2003b}.  
We take advantage of the large number of pixels in our map, and bin the
data according to the excitation temperature of the \tw\, determined as
described above.  
In each bin, we have a sufficient number of pixels that the \th\ J = 1$\rightarrow$0 line is detected with good signal to noise ratio.  
For each $T_{ex}$ bin, we then have the \tw\ excitation
temperature and the observed \tw/\th\ integrated intensity ratio.  
The data generally have the observed intensity ratio decreasing with 
increasing $T_{ex}$, from $\simeq$ 22 for $T_{ex}$ = 4.5 K to $\simeq$ 13 
for $T_{ex}$ = 2.5 K.
The free parameters are the \tw\ column density, the \h2\ density, and the 
\tw/\th\ abundance ratio.  
The latter cannot be assumed to be a fixed value (e.g. 65), due to 
the complicating presence of isotopic enhancement due to chemical 
and/or photo effects \citep[e.g.][]{watson1976,bally1982,chu1983, 
vandishoeck1988}.  
We thus consider $R$ = \tw/\th\ between 25 and 65.


\begin{deluxetable}{cccccc}
\tablewidth{0pt}
\tablecaption{\label{bestLVG_table} \tw\ Excitation Temperature Bins in Mask 1 and Best Estimates of Their Characteristics\tablenotemark{1}}
\tablehead{
\colhead{T$_{ex}$}	&\colhead{\tw/\th}				&\colhead{Number of Pixels}&
\colhead{n(H$_2$)}	&\colhead{N(\tw)/$\delta$v}			&\colhead{\tw/\th} \\
\colhead{K}		&\colhead{Observed}				&\colhead{}&
\colhead{\cc}		&\colhead{10$^{16}$ cm$^{-2}$/kms$^{-1}$}	&\colhead{Abundance Ratio}\\
} 
\startdata
4.5     &21.7   &\phm{1}32321  	&\phm{1}125  	&0.7	&30\\         
5.5     &21.7   &113923  	&\phm{1}200	&1.0	&35\\           	  
6.5     &19.6   &202328   	&\phm{1}250	&1.4	&38\\  		       	   
7.5     &16.7   &245949   	&\phm{1}280	&2.0	&40\\       		
8.5     &14.9	&211649	  	&\phm{1}325	&2.7	&42\\
9.5    	&13.9	&175431		&\phm{1}425	&3.1	&45\\
10.5   	&13.4   &122423		&\phm{1}550	&3.6	&50\\
11.5   	&13.0	&\phm{1}65428	&\phm{1}850	&3.7	&55\\
12.5   	&12.8	&\phm{1}27387	&1200 		&4.3	&65\\
\enddata
\tablenotetext{1} {The last three columns are model values.}
\end{deluxetable}
%
%

With three free parameters and only two observables, we cannot uniquely
determine the properties of the gas in mask 1.  
Rather, we compute for each $T_{ex}$ bin, a family of $R$, density and
CO column density per unit line width solutions.  If we knew
{\it a priori} the value of $R$, then we could compute a unique
density and CO column density per unit line width for each $T_{ex}$.
With no knowledge of $R$, then the values of
density and CO column density per unit line width span
a range of approximately a factor of 4, with density and CO 
column density per unit line width inversely correlated.
The family of solutions for the physical parameters of the gas show some
significant general characteristics.  
First, for higher values of $T_{ex}$, only solutions with $R\geq50$ 
fit the data.  This is encouraging as the higher excitation gas has on 
average the largest column density and we do not expect significant 
fractionation in the more shielded regions.  
On the other hand, for lower values of $T_{ex}$, values of $R$ as large as 65
are excluded, and the range of acceptable solutions gradually shifts
from $R$ $\leq$ 50 at $T_{ex}$ = 7.5 K to values of $R$ $\leq$ 30 at $T_{ex}$ = 4.5 K.  
Correspondingly, the allowable solutions for the gas density and CO
column density per unit line width decrease with decreasing excitation
temperature.  
This trend again is consistent with increasing
fractionation in the less well--shielded regions at the periphery of the 
clouds (see Liszt 2007) for a discussion of this effect in diffuse clouds).  
These regions dominate the positions found within our mask 1.
This result agrees with the behavior found in previous
observational studies \citep[e.g.][]{goldsmith1980, langer1980,
young1982, langer1989, goldsmith2005, kainulainen2006}.

It is not possible to model the mask 1 observations with a fixed value
of the {\it in situ} carbon monoxide isotopic ratio but rather require that 
the value of $R$ vary significantly with excitation temperature.  
We have chosen solutions such that $R$ varies smoothly from a value of 
65 at $T_{ex}$ $\geq$ 12.5 K to a value of 30 for $T_{ex}$ = 4.5 K.  
With this choice of $R$, we find that the gas density and
CO column density per unit line width both increase monotonically with
increasing excitation temperature.  The solutions we have chosen are 
shown in Table \ref{bestLVG_table} and in Figure \ref{bestLVG_figure}.  
We emphasize that these solutions are not unique, but depend on
our choice of $R$.
However, the general behavior of the solutions are physically plausible,
given that we expect the excitation temperature to increase as one moves
from the cloud interior to the cloud periphery.  
This suggests that binning by $T_{ex}$ is a useful approach, and gives us 
a reasonable handle on how the physical conditions vary as a
function of excitation temperature and position in the cloud.

Our assumption of 15 K for the kinetic temperature is a potential source of
error in determining the carbon monoxide column density.  To assess
this, we have carried out some calculations using a kinetic
temperature of 25 K which seems an upper limit to what one might
expect in a cloud edge in a region with modest UV intensity.  We find
that for this value of the kinetic temperature, the column
density per unit velocity gradient is approximately a factor 1.5
larger than for a kinetic temperature of 15 K, and the derived \h2\ density
is a factor of 2.5 lower, for an assumed value of $R$.  The same
trends of carbon monoxide column density and \h2\ density as a
function of $R$ are seen for the higher kinetic temperature as for the
lower.  The uncertainty resulting from the assumption of a fixed
kinetic temperature is thus of the same order as resulting from our
choosing a best value of $R$, and combining these could yield a
factor of 2 uncertainty in $N$(CO).  Observations of multiple transitions
of carbon monoxide isotopologues would provide a more accurate estimate of
the molecular column density.  However, observations of these transitions
over a region of comparable size would pose a formidable challenge for currently
available telescopes and receiver systems.

To obtain the column density for each line of sight, we utilize an analytic
fit to the relationship between the CO column density per unit line width and
the excitation temperature obtained for the set of $T_{ex}$ bins,  
$N(^{12}CO)/\delta v = (-1.473\times 10^{16}+4.672\times 10^{15}T_{ex})$.  We multiply the results by the observed FWHM \tw\ line width $\Delta v$ from the data.
The use of the LVG model introduces some uncertainty because the carbon monoxide excitation is quite subthermal, and the excitation temperature does depend on the optical depth, and is quite different for \tw\ and \th.  Nevertheless, the likely error in the trapping predicted by the LVG and other models is relatively modest compared to other uncertainties inherent in this analysis. 

%
%
\bf[!htbp]
\includegraphics[scale=0.8]{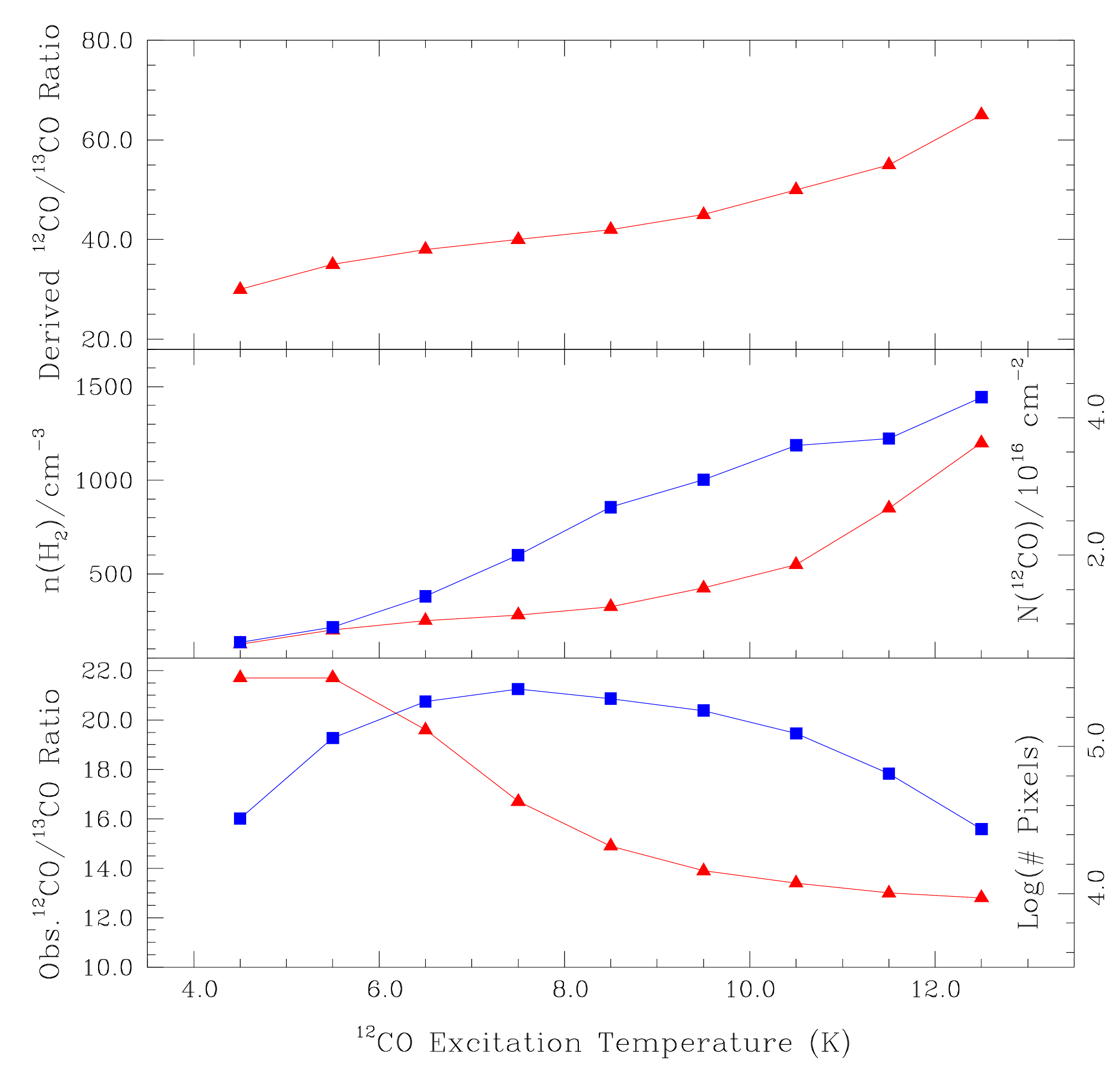}
\caption{\label{bestLVG_figure} Parameters of mask 1 pixels binned by
\tw\ excitation temperature $T_{ex}$.  The bottom panel shows the
observed \tw/\th\ intensity ratio (left hand scale; triangles), and the number of
pixels in each excitation temperature bin (right hand scale; squares).
The most common excitation temperatures are between 6 K
and 10 K.  The middle panel shows the \h2\ density (left hand scale;
triangles) and \tw\ column density assuming a line width of 1 \kms\ (right hand scale; squares). 
These are obtained from the \tw\ and \th\ intensities.  
The top panel shows the derived \tw/\th\ abundance ratio.  
The \h2\ density, \th\ column density, and the derived \tw\/\th\ ratio all increase
monotonically as a function of \tw\ excitation temperature.} \ef

%
%
\bf[!htbp]
\includegraphics[scale=0.8]{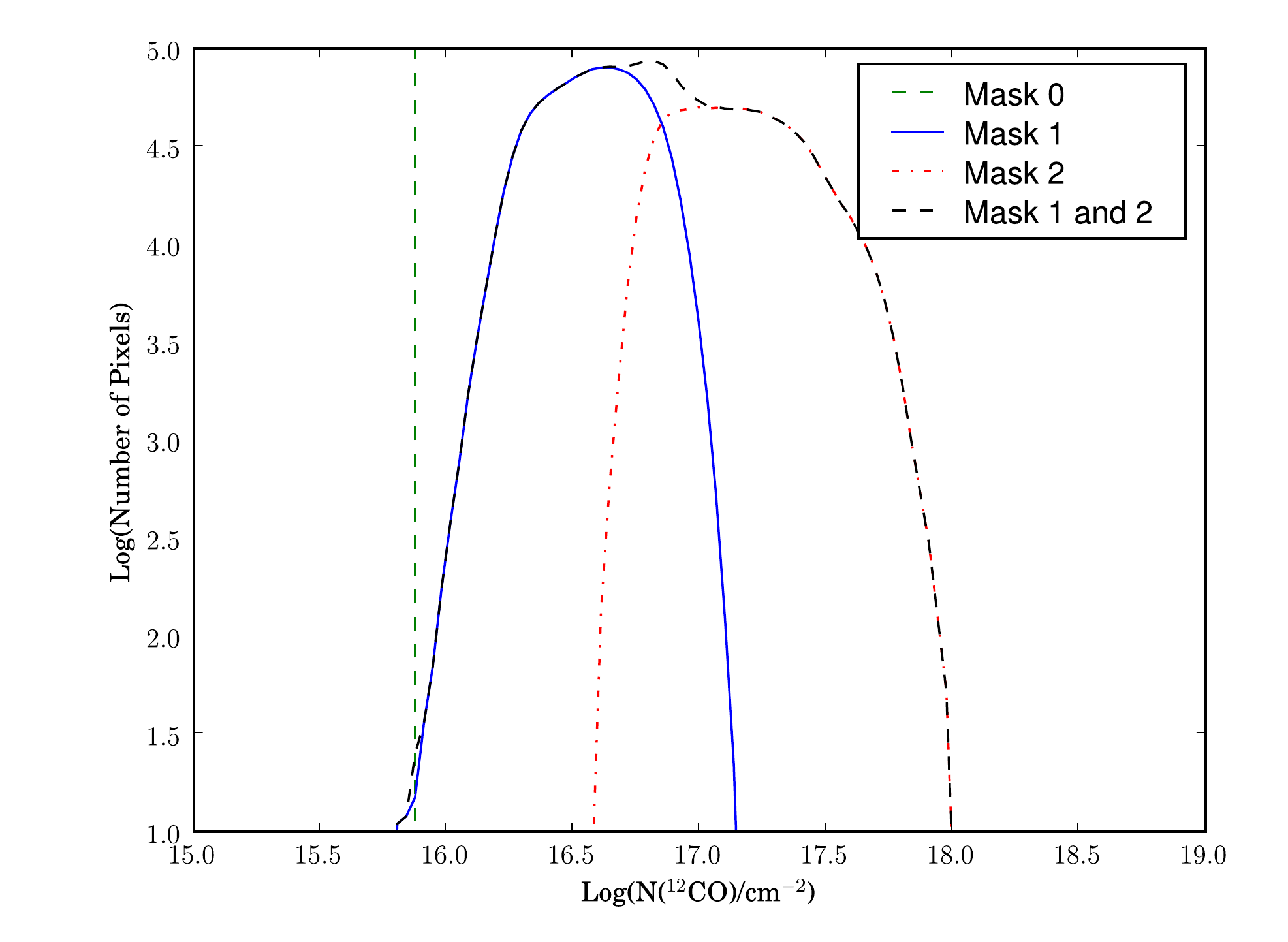}
\caption{\label{co-coldens_dist} Histogram of \tw\ column density
distributions in mask 0, 1, and 2 regions mapped in Taurus.  The mask
0 data are indicated by the vertical dashed line, which represents
 the approximately 10$^6$ pixels in this region.  The
distribution for mask 1 is shown by the solid (blue) curve, that for
mask 2 by dot--dashed (red) curve, and the combination of the two
regions by the long--dashed (black) curve.  The abscissa is the
logarithm of the \tw\ column density, defined in bins of width 0.1
dex, except for mask 0 which is a single value.  The ordinate is the
logarithm of the number of pixels in each column density bin.  The
total number of pixels included is close to 3.1 million.  } \ef

In mask 0, after averaging $\simeq$ 10$^6$ spatial pixels, we are able
to detect both isotopologues, and we thus analyze the emission for the
region as if it were a single spatial entity.  The general analysis
follows the procedure described above for mask 1.  
The fact that the integrated \tw/\th\ ratio is $\simeq$ 19 indicates that the \th\ is almost certainly optically thin. 
This is also the case for mask 1, and here as well results in the \tw\ and \th\ having quite different excitation temperatures due to the radiative trapping for the more abundant isotopologue.  

Again, we fix the kinetic temperature to be 15 K, reflecting increased heating
in regions of low extinction, and assume that the average line width is 2
\kms, similar to that observed for the low excitation gas of mask 1. 
Note that the average mask 0 spectrum (Figure \ref{3masks_spectra}) is
much broader than 2 \kms, but the large value of the line width reflects 
changes in the line center velocity over the entire region observed.
Following the trend of $R$ from mask 1, we assume this ratio to have a
value of 20.  

The mask 0 data cannot be fit satisfactorily by larger values of $R$
thus confirming that relatively strong isotopic selective effects are at work
in the low density/low column density regions of Taurus.
With these assumptions, the parameters we derive, although again not unique as
described above, are $n(H_2)$ = 75 \cc, and $N$(\tw) = 7.5$\times$10$^{15}$ \c2.  
The carbon monoxide excitation in this region is evidently highly
subthermal, consistent with the low derived H$_2$ density and the
modest \tw\ optical depth.  
This very low value for the density of the mask 0 region gives a reasonably
low column density for the extended component of the gas in Taurus.  
Taking a representative dimension for mask 0
of 1.5$\times$10$^{19}$ \c2, we obtain $N(H_2)$ = 1.1$\times$10$^{21}$ \c2.
This corresponds to A$_v$ $\simeq$ 1 for the extended component of the cloud,
consistent with that determined from stellar reddening \citep{cernicharo1987}.

The spatial distribution of column densities from the three mask
regions is shown in Figure \ref{co-coldens_dist}.
The column density for mask 0 is a single value $<N(^{12}CO)>$ =
7.5$\times$10$^{15}$ \c2 as given above. The column density distribution
in the mask 1 region is a relatively symmetric, fairly Gaussian
distribution with a mean value $<N(^{12}CO)>$ = 3.6$\times$10$^{16}$
\c2.  The column density distribution in the mask 2 region is
flat--topped with a mean value $<N(^{12}CO)>$ = 1.3$\times$10$^{17}$
\c2.

The distribution of carbon monoxide in the Taurus region is shown in
Figure \ref{co_coldens_map}.  This figure dramatically illustrates the
complexity of the molecular gas distribution.  The
impression given is quite different from that of studies with low
angular resolution, in that instead of an ensemble of ``relaxed",
fairly smooth condensations one sees a great deal of highly
filamentary structure, a strong suggestion of cavities and surrounding
regions with enhanced column densities.  The large size of the region
covered also suggests relationships between the different portions of
the Taurus molecular region.  The most striking of these points will
be addressed briefly later in this paper.

%
%
\bf[!htbp]
\includegraphics[scale=0.9]{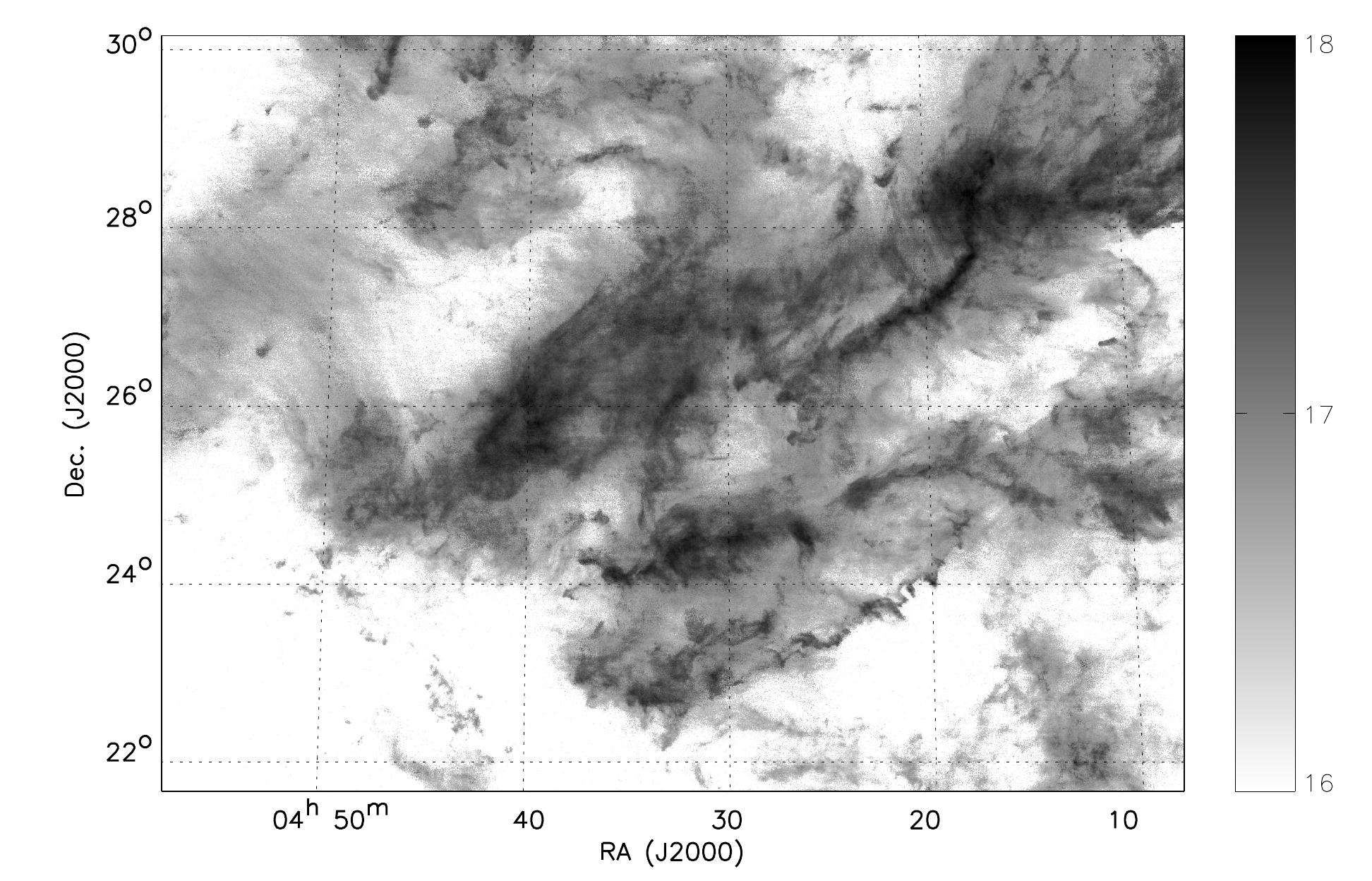}
\caption{\label{co_coldens_map} Distribution of carbon monoxide in
Taurus region expressed as the logarithm of the \tw\ column density (cm$^{-2}$) in
mask 1 and mask 2 regions.  The scale is indicated by the bar on the right hand side.
The \tw\ column density in the mask 0 region (indicated by light grey density) 
is 7.5$\times$10$^{15}$ \c2, as discussed in the text.  The maximum \tw\ column 
density is a factor of $\simeq$ 100 larger.}  
\ef

\subsubsection{Molecular Hydrogen Column Density and Mass}

Most studies of molecular regions using carbon monoxide have
emphasized regions in which the column density is sufficiently large
that dust shielding plus self--shielding result in an ``asymptotic''
\tw\ abundance between 0.9$\times$10$^{-4}$ and 3.0$\times$10$^{-4}$
relative to \h2\ \citep[see e.g.][]{frerking1982, lacy1994}.  In our
study of Taurus, only the mask 2 region is plausibly consistent with
this assumption.  The remainder of the cloud is characterized by lower
densities and column densities, and the fractional abundance of carbon
monoxide must be regarded as being significantly uncertain and likely
to be dependent on the extinction.

There is considerable value in trying to make a self--consistent model for the
carbon monoxide as a tracer of total molecular (H$_2$) column density.
To this end, we have used the theoretical modeling by
\cite{vandishoeck1988}.  We have utilized the curve for $I_{UV}$ = 1.0
(in units of Habings), carbon depletion $\delta_C$ = 0.1,
and models T1--T6, which correspond to temperature range 40 K to 15 K
and n$_H$ = 500 \cc\ to 1000 \cc\ throughout the model slab being
considered.  We have used a polynomial fit to the data from the appropriate curve in
Figure 8 of \cite{vandishoeck1988} for the relationship between CO and
H$_2$ column densities.  
This value of carbon depletion is recommended by \cite{vandishoeck1988} as
agreeing with the available Taurus data.
We also note that the carbon monoxide fractional abundance as given
by these models of \cite{vandishoeck1988} agrees well at low column densities 
with the UV measurements of \cite{sonnentrucker2007} and \cite{burgh2007}. 

The lower CO lines in absorption from diffuse clouds lying in front of millimeter continuum sources have been observed by \cite{liszt1998}. The clouds, analyzed by \cite{liszt2007} have a range of \h2\ column density (determined by UV absorption; Federman et al. 1994) which extends from 5$\times$10$^{20}$ \c2\ to just above 10$^{21}$ \c2, and thus includes our mask 0 (and very low end of mask 1) results.  
While there is considerable scatter among various clouds having the same hydrogen column
density, the best fit relationship gives $X$(\tw) = 5$\times$10$^{-6}$ for $N$(\h2) = 10$^{21}$ \c2.  This is quite close to our results and again reinforces the general applicability of a reduced carbon monoxide fractional abundance for low extinction cloud material.  The specific parameters we have adopted have been chosen, in addition to being consistent with the measurements of low column density diffuse clouds,  to give good agreement at high column densities with the mm emission measurements of \cite{bachiller1986}, \cite{cernicharo1987}, and \cite{alves1999}.

The strong dependence of CO column density on H$_2$ column density
reflects the onset of self--shielding when N(CO) reaches $\simeq$10$^{15}$
\c2.  This produces a  rapidly increasing CO fractional
abundance as a function of H$_2$ column density in the range covered
by the mask 0 and mask 1 regions of our study, and a gradual leveling out
of N(CO)/N(\h2) in mask 2. 
The most significant difference is that using this approach we find that the
low CO column densities correspond to considerably larger \h2\ column densities
than would be found if a constant fractional abundance of CO were adopted. 
We convert our CO distribution to a molecular hydrogen distribution using the 
nonlinear relationship, and the result is given in histogram form in Figure \ref{h2-dist-image}.

%
%
\bf[!htbp]
\includegraphics[scale=0.8]{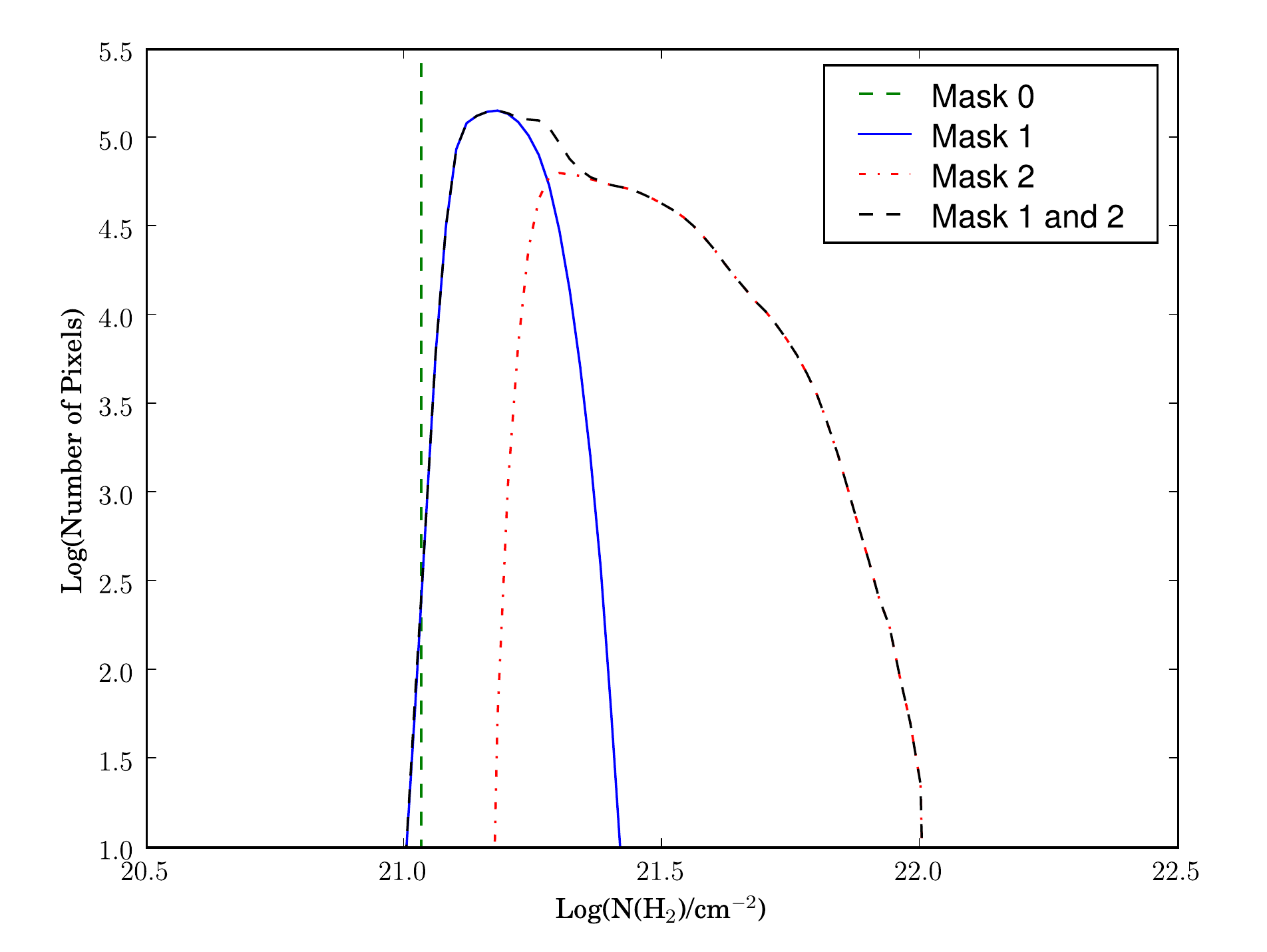}
\caption{\label{h2-dist-image} Histogram of H$_2$ column density
distribution in Taurus.  The distribution of column density in the mask 0 region is given by the 
vertical (dashed) line, in the mask 1 region by the solid (blue) curve, in the 
mask 2 region by the dot--dashed (red) curve, and in the combination of the two regions
by the long--dashed (black) curve.}  
\ef

When compared to Figure \ref{co-coldens_dist}, it is evident that the
varying fractional abundance has resulted in a significant compression
in converting the carbon monoxide to \h2\ column densities.  The
drop in X(CO) in regions of lower extinction and lower density means
that the relatively weak emission that we observe there implies a greater H$_2$
column density than would be derived assuming a constant fractional
abundance.  Taking mask 0 as an example, the CO column density of
7.5$\times$10$^{15}$ \c2, with fractional abundance 7.0$\times$10$^{-6}$
corresponds to an H$_2$ column density equal to
1.1$\times$10$^{21}$ \c2\ using the variable fractional abundance,
more than an order of magnitude larger than would be obtained using the
canonical high--extinction fractional abundance of 10$^{-4}$.

This suggests that the majority of the area within the Taurus 
molecular cloud complex has a visual extinction from
molecular hydrogen on the order of 1 magnitude.
This is consistent with the hydrogen column density of mask 1 discussed in
the previous section, as well as with the ''halo'' component of the HCL2 region
discussed by \cite{cernicharo1987}.  There is certainly a
high column density tail which reaches 10$^{22}$ \c2, but this
includes only a very small fraction of the cloud area and mass.  While
\th\ is not the ideal tracer of the densest component of the cloud, this
study makes it clear that only about 10$^{-3}$ of the pixels with \tw\
detectable have $A_v$ $\geq$ 5.

Despite the relatively low density in mask 0 and mask 1 regions, the time scale to arrive at the the low fractional abundance of carbon monoxide found there is quite modest.
Using the expression from Section 4.1 of \cite{liszt2007}, we find that if we start with $X(e)$ = 10$^{-5}$ and $n$(\h2) = 100 \cc, the characteristic time to reach $X$(CO) = 10$^{-5}$ is only $\sim$ 10$^5$ yr.  This is consistent with results obtained using explicit time--dependent models with CO formation and destruction by E. Bergin (private communication).  
Thus, whatever the history of the diffuse surroundings of dense clouds, the low but significant abundance of carbon monoxide found there appears entirely plausible.

We show the spatial distribution of \h2\ column density in the Taurus region
mapped in Figure \ref{coldensmap}.  The contributions of individual
pixels in mask 1 and mask 2 are included.
Approximately 50 percent of the total molecular mass of the region is
in directions in which \th\ cannot readily be detected in an
individual map pixel.  From the masses in each mask region, we compute
the total mass of the region of Taurus mapped in the present study.
The results (including correction for He and heavy elements) are given in Table \ref{mask_mass}.  For mask 0, we have
considered the entire area it comprises to be characterized by the
single set of conditions derived in the previous subsection, while the
contribution of mask 3 has been neglected.

Table \ref{mask_mass} shows that assuming the physically plausible variable fractional
abundance of carbon monoxide gives a total mass of the region a factor approximately
2.5 times larger than that obtained using a uniform high abundance characteristic of
well--shielded regions.  We also see that the contributions from the low column density
mask 0 and mask 1` regions are considerably enhanced and that their contribution 
to the total mass is no longer negligible as would be the case if a constant fractional
abundance obtained.

%
%
\bf[!htbp]
\includegraphics[scale=0.9]{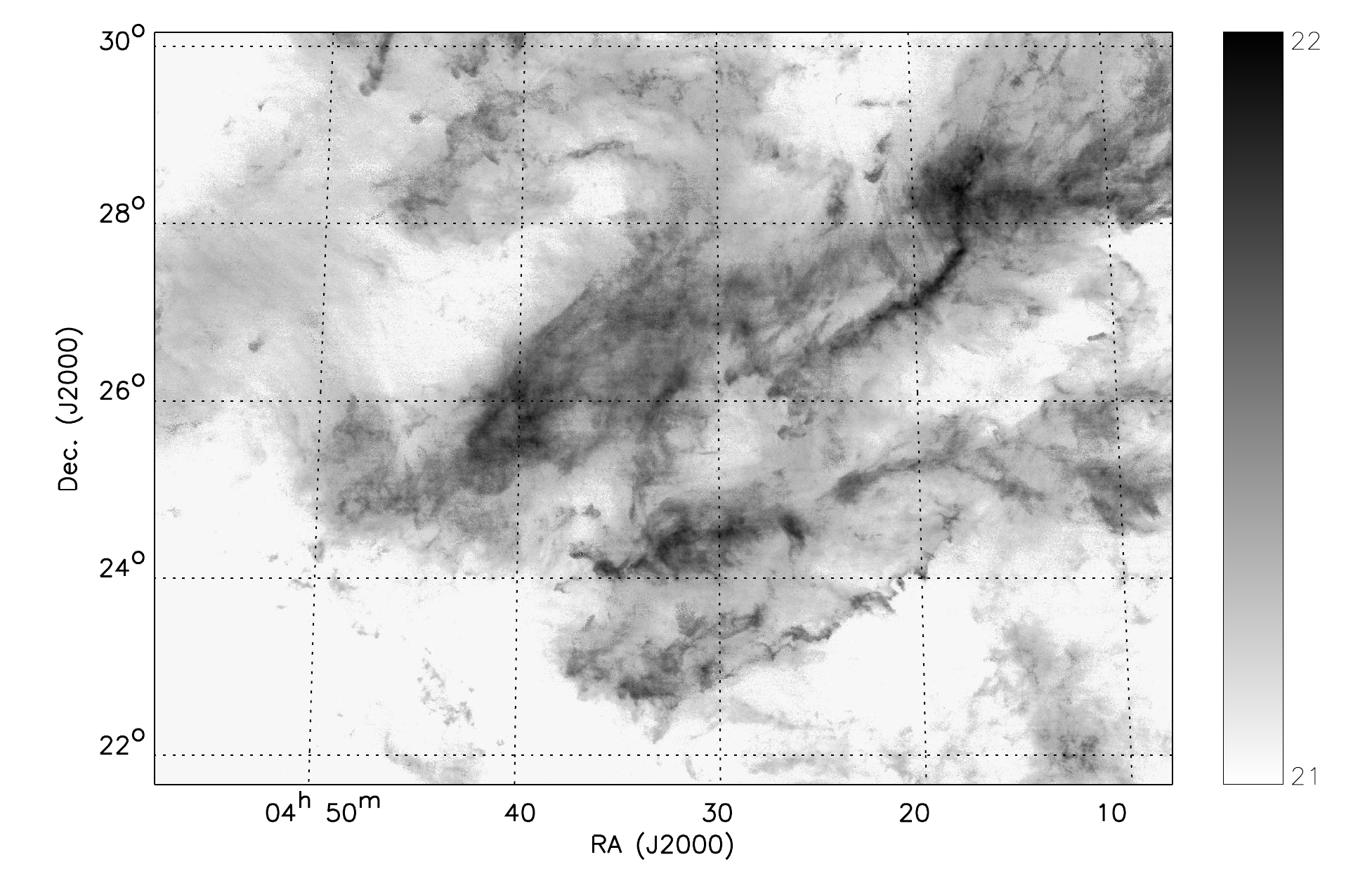}
\caption{\label{coldensmap} Image showing the molecular hydrogen
column density distribution derived from mask 0, mask 1, and mask 2
regions in Taurus.  The scale is indicated by the bar on the
right, expressed as the logarithm of the derived \h2\ column density
in cm$^{-2}$.  } \ef


\begin{deluxetable}{ccc}
\tablewidth{0pt}
\tablecaption{\label{mask_mass} Mass of Region in Taurus Mapped}
\tablehead{\colhead{Mask Region}	&\colhead{Mass (10$^3$ \Ms)}		&\colhead{}\\
		\colhead{}		&\colhead{a}				&\colhead{b}\\
}			
\startdata
0	&0.1		&\phm{1}4.1 \\
1	&1.7		&\phm{2}7.7 \\
2	&7.8		&11.8\\
{}	&{}		&{}  \\
Total	&9.6		&23.6\\
\enddata
\tablenotetext{a}{Using constant H$_2$/CO ratio equal to 2$\times$10$^4$}
\tablenotetext{b}{Using H$_2$/CO ratio with I(UV) = 1.0 and $\delta_C$ = 0.1 from \cite{vandishoeck1988}}
\end{deluxetable}

\subsection{Cloud Structure}

Valuable insight into the structure of the cloud can be obtained by
examining the cumulative distribution of cloud mass and area as a
function of column density.  This information in shown in Figure
\ref{cum_mass_area}.
Our survey focused on the region of the
Taurus molecular cloud known to have most prominent high density
regions with exceptional chemical diversity
\citep[TMC-1;][]{pratap1997} and prominent star formation
(e.g. L1495).  
Nevertheless, we see that half of the cloud's mass is in material with
N(H$_2$) less than 2.1$\times$10$^{21}$ \c2.  Only about 5\% of the
cloud's mass occurs at \h2\ column densities above 5$\times$10$^{21}$ \c2,
or visual extinction greater than 5. 
The column density we derive may be modestly underestimated due to 
incomplete correction for saturation  in our \th\ observations for large 
column densities, and as a consequence of molecular depletion at high densities, but even
together these effects are unlikely to increase this fraction by a
factor of 2 \citep[see e.g.][]{alves1999}.  The fraction of the cloud
area with N(H$_2$) $\geq$ 5$\times$10$^{21}$ \c2\ is only 0.02.

%
%
\bf[!htbp]
\includegraphics[scale=0.7]{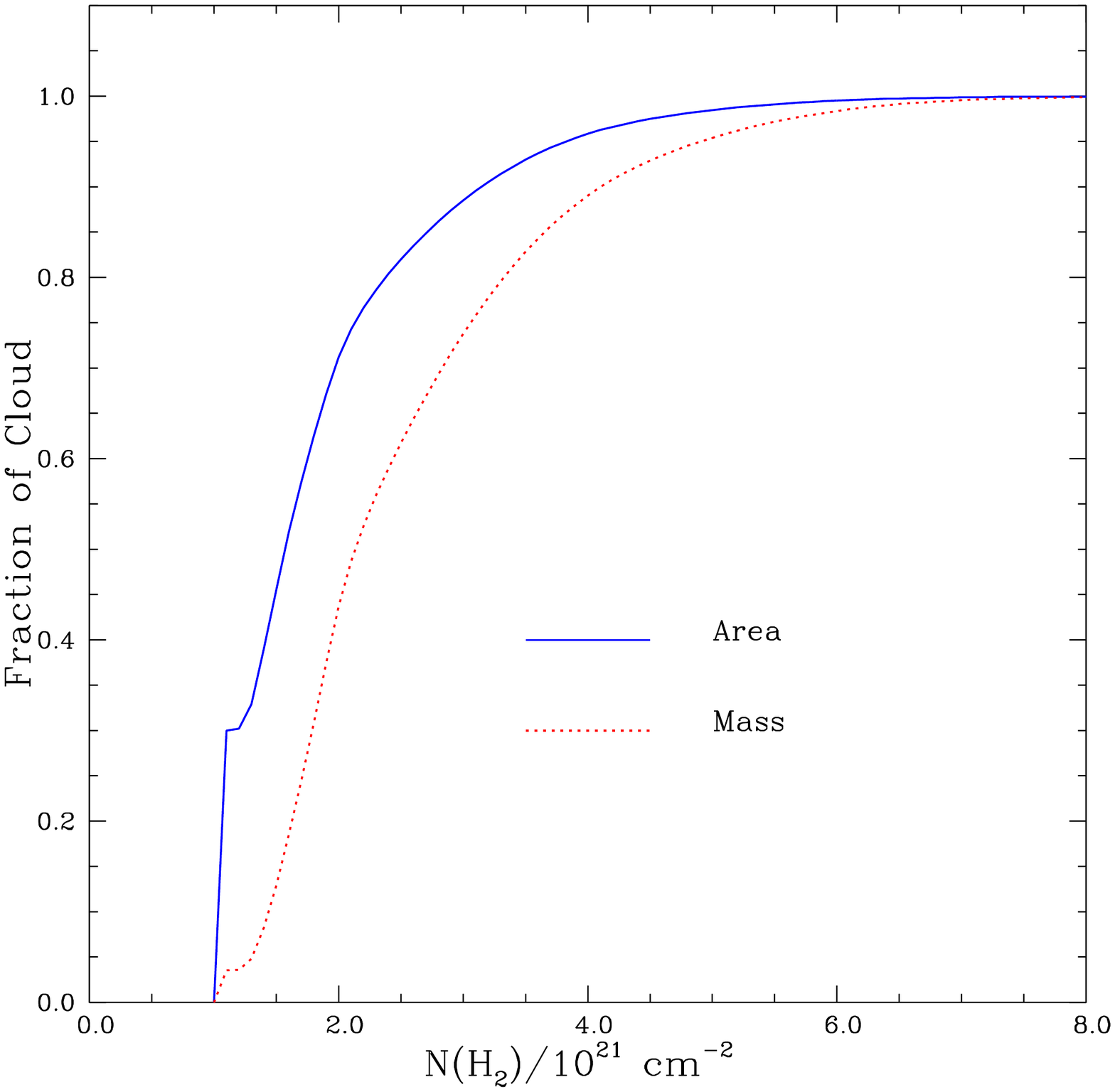}
\caption{\label{cum_mass_area} Cumulative fraction of Taurus as a
function of molecular hydrogen column density.  The solid (blue) curve
gives the fraction of the area characterized by column density less
than specified value, and the dotted (red) curve gives the fraction of
the mass similarly characterized.  Half the area mapped has column
density below 1.6$\times$10$^{21}$ \c2, and half the mass of the cloud
is included in regions having column density below
2.1$\times$10$^{21}$ \c2.}  
\ef

Another view of the mass distribution can be obtained by attempting to
dissect the cloud by extracting the well--recognized high column
density regions from the remainder of the gas.  In Figure
\ref{roi_boundaries} we show the division into eight regions, which
together include approximately 25\% of the area of the map.  We have
generally followed the region limits and designations given in Fig. 3
of \citet{onishi1996}.

%
%
\bf[!htbp]
\includegraphics{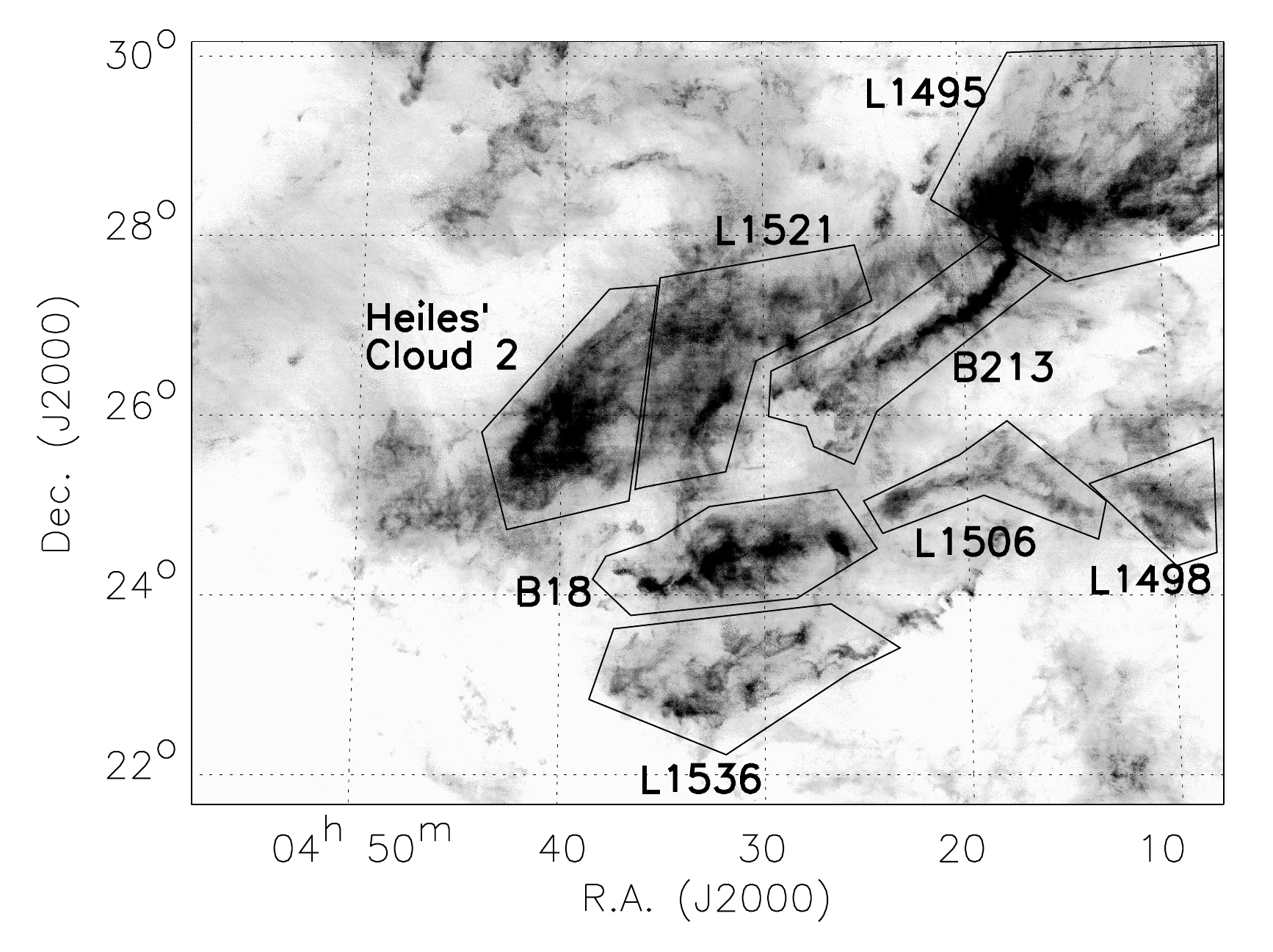}
\caption{\label{roi_boundaries} Image showing H$_2$ column density 
overlaid with the eight well-known regions of high column
density as designated by Onishi et al.\ (1996).  These regions define
the masses and areas given in Table \ref{roi_masses}.}  
\ef

We give the mass of each of these regions in Table \ref{roi_masses}.
The total mass contained in these regions, 9807 \Ms, is 42\% of the total mass
included in the region we have studied, and their combined area is 21\% of that
of the region we have mapped.  However, since we have made an unbiased
map of \th\ rather than a map restricted to regions of strong
intensity \citep[as][did in their \th\ survey]{mizuno1995}, we
include somewhat larger areas.  The masses we derive for L1495/B213
and for B18 are approximately a factor of 3 larger than those obtained
by \cite{mizuno1995}, and that for HCl2 is a factor of 2 larger.  It
is evident that a large fraction of the mass even within the
boundaries shown in Figure \ref{roi_boundaries} is in relatively
low--density gas.


\begin{deluxetable}{lcc}
\tablewidth{0pt}
\tablecaption{\label{roi_masses} Mass of High--Density Regions in Taurus\tablenotemark{a}}
\tablehead{	\colhead{Region}	&\colhead{Mass}\tablenotemark{b}	&\colhead{Area}\\
		\colhead{}		&\colhead{\Ms}				&\colhead{pc$^2$}\\}
		
\startdata
L1495	&2616		&\phm{1}31.7\\
B213	&1095		&\phm{1}13.7\\
L1521	&1584		&\phm{1}17.6\\
HCl2	&1513		&\phm{1}15.8\\
L1498	&\phm{1}373	&\phm{11}5.7\\
L1506	&\phm{1}491	&\phm{11}7.7\\
B18	&1157		&\phm{1}14.5\\
L1536	&\phm{1}978	&\phm{1}16.6\\
{}	&{}		&{}\\
Total	&9807		&123.3\\
\enddata
\tablenotetext{a}{Regions defined in Figure \ref{roi_boundaries}}
\tablenotetext{b}{Includes correction for He}
\end{deluxetable}

Having a well--sampled \tw\ map of a large region and a mass determination allows us to examine the application of a CO luminosity to mass conversion factor \citep{dickman1986} to Taurus.
In Table \ref{mass_lum} we show the results for the different mask regions and the
total.  The entries in the third column are obtained using a conversion factor 
M(\Ms) = 4.1L$_{CO}$(K km s$^{-1}$ pc$^2$).  
This value is obtained using the Egret $\gamma$--ray data \citep{strong1996}, and a factor 1.36 for the total mass per H$_2$ molecule (including He and metals) in the gas.
For mask 0, the CO luminosity drastically underestimates the mass, due to highly subthermal excitation of the CO and its modest optical depth.  For the denser regions, the agreement is much better.
The surprisingly close agreement for the complete Taurus region may, to a certain extent, be fortuitous, but it suggests that use of the \tw\ luminosity to derive total mass of molecular regions does appear to work reasonably well for regions with only low--mass young stars, as well as for regions with young high--mass stars.

\begin{deluxetable}{lccc}
\tablewidth{0pt}
\tablecaption{\label{mass_lum} Comparison of Masses Determined from \tw\ and \th\ With Those Derived from CO Luminosity}
\tablehead{\colhead{Region}	&\colhead{Mass from} 	&\colhead{\tw\ Luminosity}	  &\colhead{Mass from}\\
	   \colhead{ }		&\colhead{\tw\ and \th} &\colhead{}			  &\colhead{\tw\ Luminosity}\\
	   \colhead{ }		&\colhead{(\Ms)}	&\colhead{(K km s$^{-1}$ pc$^{2}$)}&\colhead{(\Ms)}\\
}
\startdata
mask 0	&\phm{1}4081	&\phm{1}193	&\phm{22}791\\
mask 1	&\phm{1}7699	&2052		&\phm{2}8413\\	
mask 2	&11752		&3305		&13550\\
\hline
Total	&23532		&5550		&22754\\
\enddata
\end{deluxetable}

\section{LARGE SCALE KINEMATICS OF THE MOLECULAR GAS}
\label{kinematics}

Previous studies have revealed a variety of motions on different
scales within the Taurus complex.  These include velocity gradients
along individual filaments possibly indicative of rotation, along with
a systematic East--West velocity difference as one moves across the
region.  In Figure \ref{gas_kin} we show a color--coded image of the
integrated intensities in three velocity intervals for the two isotopologues.
There is a great deal of structure seen even in this relatively crude representation of
the velocity field.  
Certain regions, and particularly the edges of particular regions, show up as having
significantly shifted velocities relative to the surrounding gas.  

This coarsely divided integrated intensity does not give the full measure of the complexity of the \th\ and \tw\ line profiles in Taurus.  
An indication of this can be seen in Fig. 20 of \cite{narayanan2007}, in which it
is evident that in general the regions with multiply--peaked lines exhibit this
characteristic in both \tw\ and \th.  
Since the visibility of the multiple peaks is approximately equal in the two isotopologues,
it is unlikely to be a result of self--absorption, but rather an indication of multiple,
kinematically distinct components.  
These are most prominent in several regions of Taurus, notably the western part of B18,
north of L1521, in B213 and west thereof, and in the southern part of Heiles' Cloud 2.  
This indicates that some regions are characterized by a considerably greater degree of
velocity multiplicity along lines of sight. 
There does not appear to be any correlation of this characteristic with e.g. star formation.

%
%
\bf[!htbp]
\includegraphics{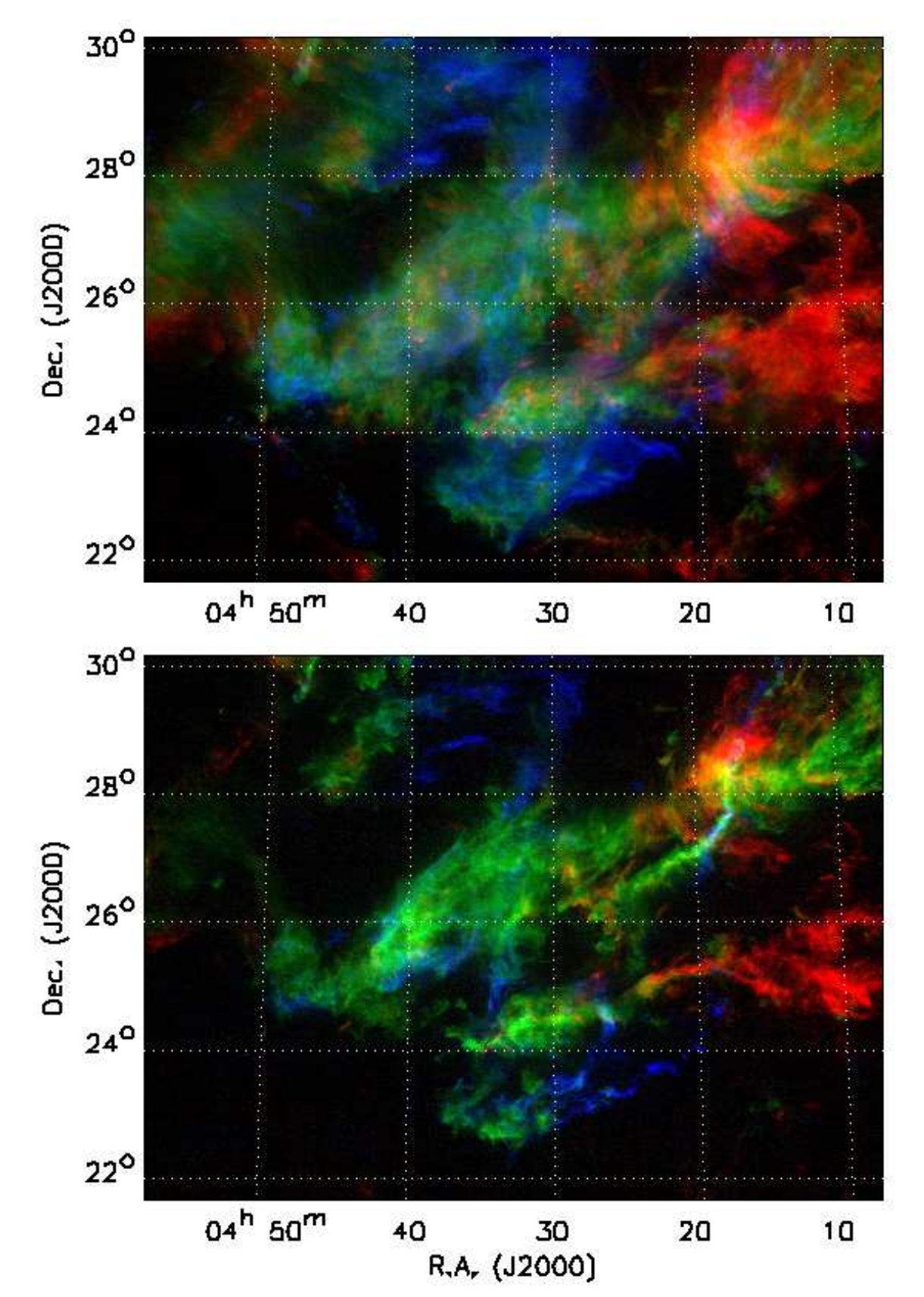}
\caption{\label{gas_kin} Color--coded image of the integrated intensities in three velocity intervals of carbon monoxide in Taurus, with emission at velocities between 3 \kms\ and 5 \kms\ coded blue, between 5 \kms\ and 7 \kms\ coded green, and between 7 \kms\ and
9 \kms\ coded red.  Upper panel: \tw\ with blue-- and red--coded
intervals scaled to 7.5 \kks\ and the green--coded interval scaled to
15 \kks.  Lower panel: \th\ with blue-- and red--coded intervals
scaled to 5.0 \kks\ and the green--coded interval scaled to 2.5 \kks.
} 
\ef
%
%
\section{MOLECULAR GAS AND THE MAGNETIC FIELD}
\label{magnetic}

The Taurus Molecular Cloud has long been a target for
investigations of the interstellar magnetic field and its role within
the dynamics of the molecular gas component \citep{moneti1984,
heyer1987, heyer1988, goodman1992, troland1996, crutcher2000}.  Many
of these studies have compared the distribution of gas and dust with
respect to the magnetic field geometry inferred from optical
polarization measurements of background stars.  The relationship of
the cloud geometry to the magnetic field morphology is an essential
aspect of models that have been developed for the formation of Taurus
\citep{gomez1992, ballesteros1999}.  These have hypothesized an
initial alignment of a more diffuse cloud with the Galactic magnetic
field as part of the initial conditions for formation of the dense
cloud, with the gas streaming along magnetic field lines.

Observationally, at intermediate scales ($\sim$1 pc), the situation
has become more complex.  In particular, toward the western end of the
Taurus cloud, the long axis of the L1506 filament is oriented along
the field in contrast to alignments of Heiles' Cloud 2 and the
B216 and B217 filaments for which the field is essentially perpendicular
to the axis of the filaments \citep{goodman1992}.  
Note that the latter structure is denoted B213 in Figure \ref{roi_boundaries}.
From this departure from rigorous alignment, \citet{goodman1992} conclude that either the
magnetic field does not dominate the cloud structure at these scales
and densities, or that the optical polarization measurements probe a
volume that is spatially distinct from the dense
filaments. \citet{goodman1992} demonstrate that polarization by
selective absorption at optical and infrared wavelengths is produced
by dust grains within the outer, low column density envelopes of the
molecular clouds and provides little or no information on the magnetic
field direction within the high density filaments.

The \tw\ and \th\ data presented in this study afford an opportunity
to extend these comparisons to lower column densities than these
previous investigations.  We have used the data assembled by
\citet{heiles2000}, taken from other sources, and superimposed this on a figure showing the
integrated intensities of \tw\ and \th.  Figure \ref{CO+B} shows the results.  
%
%
\bf
\includegraphics[scale=0.7]{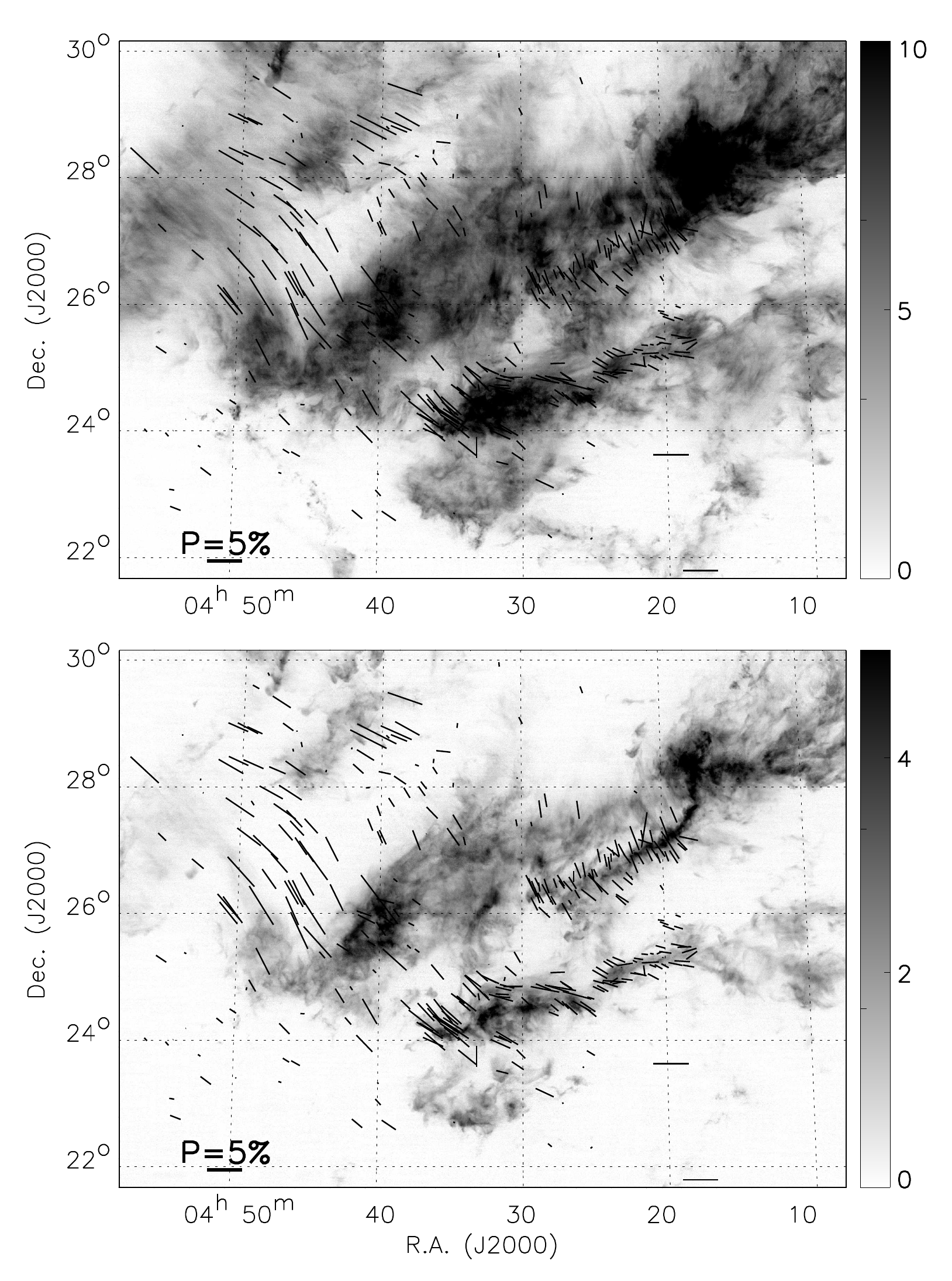}
\caption{\label{CO+B} Upper panel:  magnetic field direction in Taurus
superimposed on the \tw\ antenna temperature distribution integrated over the 
5 \kms\ to 8 \kms\ velocity range, chosen to emphasize striations and other fine structure.  The line
segments indicate the direction of the magnetic field derived from
observations of absorption by polarized dust grains; their length is
proportional to the fractional polarization.  
The horizontal bar at lower left in each panel indicates 5\% fractional polarization.
Lower panel:  magnetic field direction superimposed on the distribution of \th\ antenna
temperature integrated over the same velocity interval.
The integrated intensities for each panel in \kks\ are indicated by the bar at the right.}
\ef
This figure highlights the relationship between the field direction
and the morphology of the dense filaments of gas discussed in the
references given above.

We can use the \tw\ emission to probe the relationship between the
lower column density portions of Taurus and the magnetic field.  This
comparison is shown in the top half of Figure \ref{CO+B}.  Within the
faint, low surface brightness \tw\ emission, we see marked striations,
which are discussed in more detail in \S \ref{striations}.  Remarkably, these features
within the Taurus Cloud follow the local orientation of the magnetic
field even as the polarization angles vary from a mean of 53 degrees
within the northeast corner of the surveyed area, to 81 degrees within
the southwest corner.  The alignment of these faint features points
toward a strong coupling of the gas with the interstellar magnetic
field.  Such strong coupling may be expected in these low column
density regions that are more exposed to the ambient, UV radiation
field, which maintains a higher degree of ionization.

The origin of these threadlike features and the mechanism whereby they
are aligned with the magnetic field are not established, but we can
speculate on several processes that may be responsible.  The channel
maps of the molecular line emission identify regions of
systematic motions over scales from the resolution limit up to 30\arcmin\ to 60\arcmin.
If the magnetic field is well coupled to the neutral gas by frequent
ion-neutral collisions but the magnetic energy is small with respect
to the kinetic energy of the gas, then the field can be carried by
these large scale flows within the cloud.  Correspondingly, the field
lines would be stretched along the direction of the flow.
Alternatively, the narrow emission threads may arise from successive
compressions and rarefactions of the gas and magnetic field produced
by magnetosonic waves that propagate perpendicular to the field.
Within the subthermally excited regime, which likely prevails within
these regions of low surface brightness, these column density
perturbations would produce corresponding variations in the \tw\
intensity.
%
%
\section{MOLECULAR GAS AND YOUNG STARS IN TAURUS}
\label{youngstars} 

The distribution of young stellar objects with respect to the 
molecular gas may offer valuable insights to the formation of stars
within a dense interstellar cloud.  For comparison with our molecular
images, we adopted the set of pre-main sequence stars in the Taurus
regions from S. Kenyon (2007 private communication, to be published in
2008).  This list is comprised of data from many surveys in optical
and infrared wavebands
\footnote{We obtain essentially the same results using the data
compiled by F. Palla, which was also provided to us as a private
communication.}.  The pre--main sequence stars are divided into
three populations according to their colors.  If the R-K magnitude is
larger than eight, the star is categorized as likely to be a Class I
or younger source.  If R-K is smaller than eight, the source is likely
to be a T-Tauri star.  If the source is not detected in either R or K, it is 
is likely to be extended/nebulous, in which case it is probably
still a protostar, younger than a T-Tauri star.  In the region covered
by our map, there are a total of 230 stars, 18 of which are Class I or
younger, 44 are extended, and 168 are likely to be T-Tauri stars.  The
stars are shown overlaid on the distribution of the \h2\ column
density in Figure \ref{taurus-stars}.  The distribution of pre--main sequence 
stars generally follows that of the dense gas, although a many of 
the stars in the older category are located in regions with only diffuse gas emission.
As noted by \cite{hartmann2002}, the young stars are grouped in
three nearly parallel bands that are associated with Heiles' Cloud
2/L1521/B213/L1495, B18/L1506 and L1536.

%
\bf[!htbp]
\includegraphics[scale=0.9]{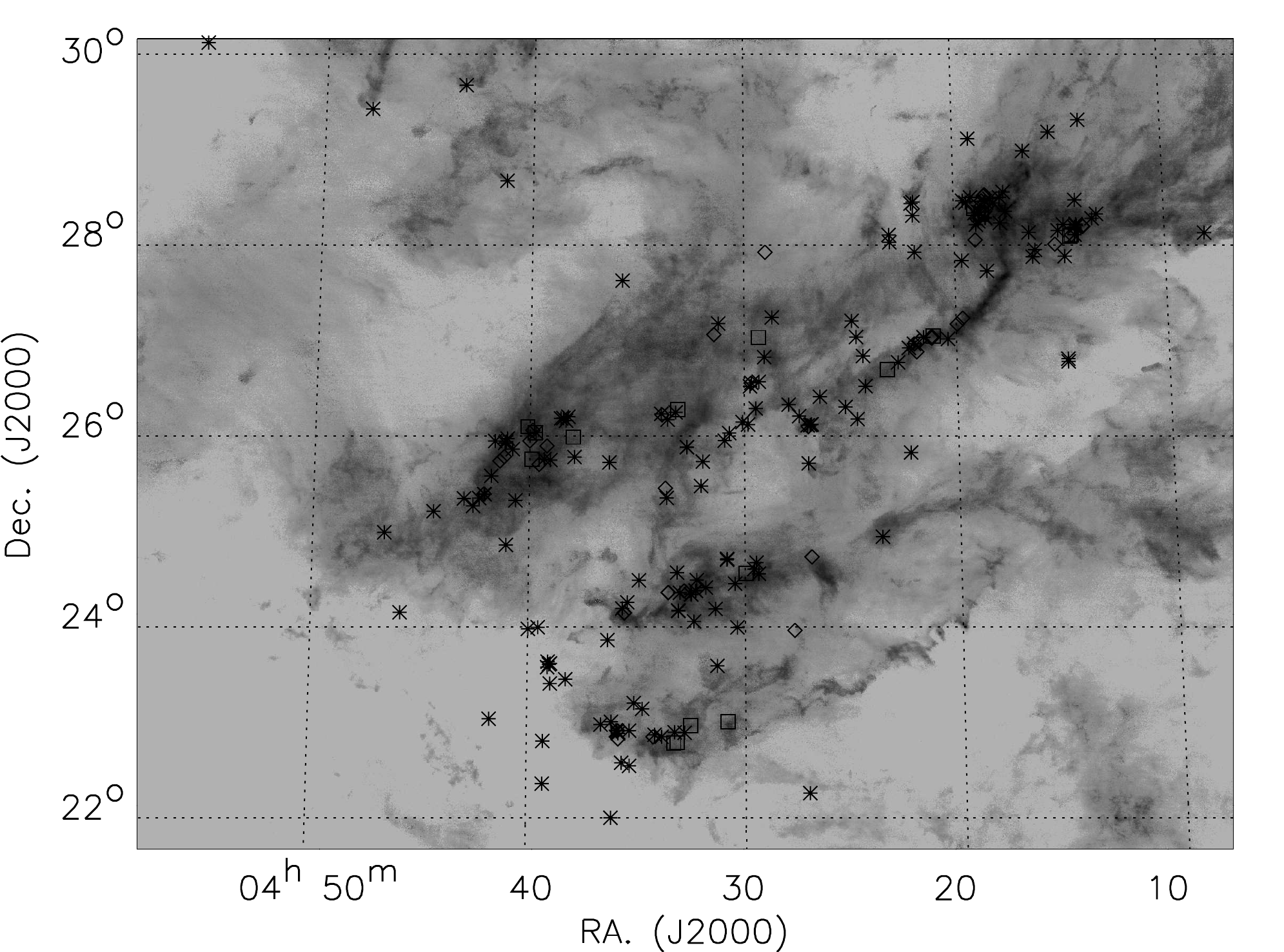}
\caption{\label{taurus-stars} Locations of young stars in Taurus
superimposed on map of the \h2\ column density.  The stellar positions are
from \cite{kenyon2007}.  The diamonds indicate diffuse or extended
sources (of which there are 44 in the region mapped), the squares
indicate Class I or younger stars (18), and the asterisks indicate
T-Tauri stars (168).  It is evident that the diffuse and
younger sources are almost without exception coincident with regions of
relatively large column density, while the older stars show a much
larger probability of being found in regions of lower column density.
} 
\ef

The relationship between \h2\ column density and stellar population is
examined further in Fig.~\ref{h2_stars}.
%
%
\bf[!htpb]
   \includegraphics[scale=0.8]{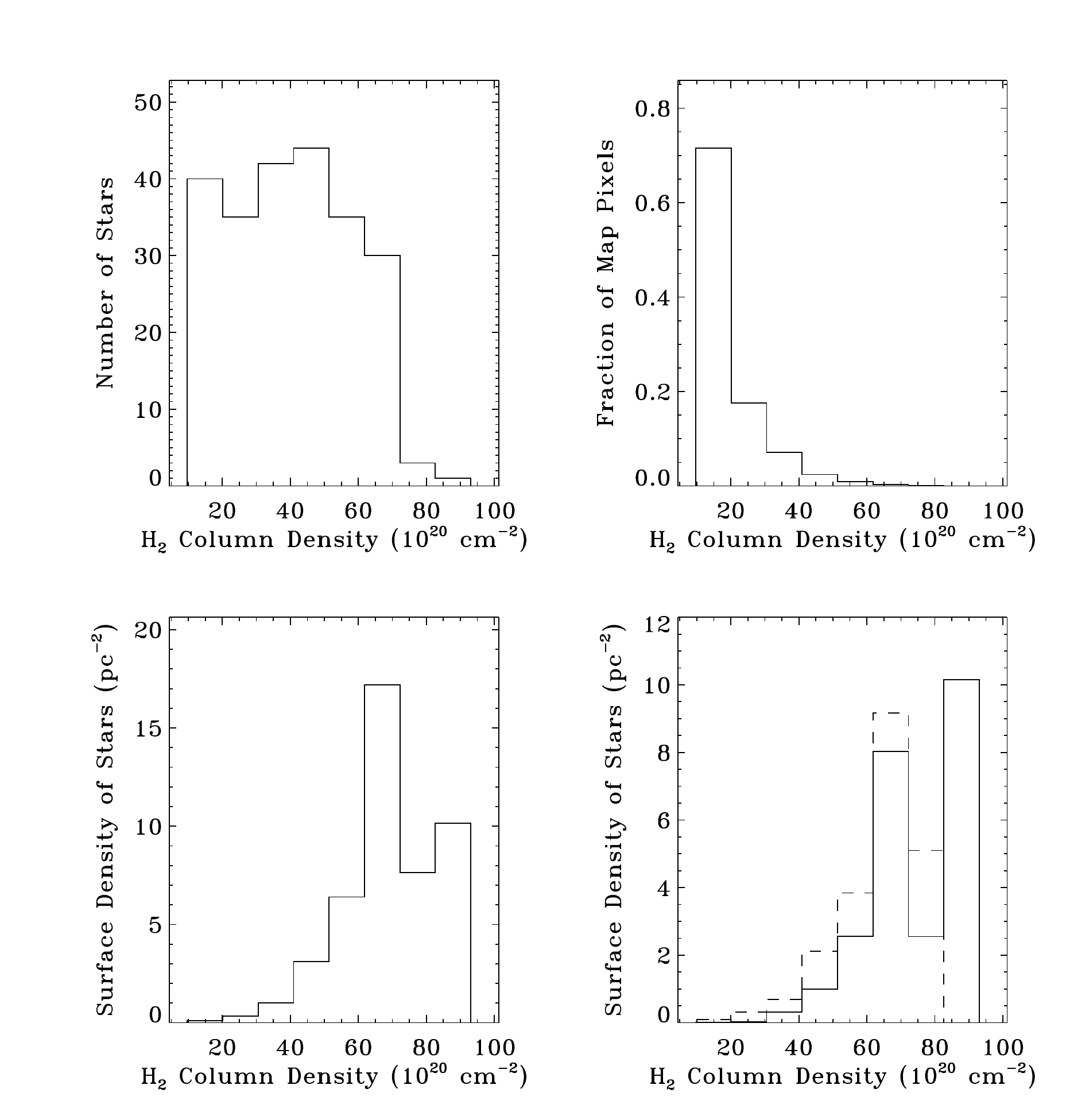}
    \caption{\label{h2_stars} Upper left: number of stars as a
    function of column density. The same bins of column density are
    used in all panels. Upper right: fraction of pixels in each column
    density bin. Lower left: surface density of stars obtained by dividing the                   number of stars in each bin by the area of the map corresponding to each bin. 
    Lower right: surface density of stars in each bin for
    likely class I and diffuse sources (solid line) and likely T-Tauri
    stars (dashed line). The division of sources is discussed in the
    text.}
    
\ef
Roughly equal number of stars can be found in each of the column
density bins spanning the range from 0 to $6.5\times10^{21}$ \c2\
(upper-left panel).  Although the number of stars drops towards
higher column density regions, such direct examination of the
distribution of stars is somewhat misleading inasmuch as our map
includes a substantial area with very weak or no carbon monoxide
emission, as shown in the upper right panel of Figure \ref{h2_stars}.  
The surface density of stars versus column density is plotted in the 
lower left panel.  A significant jump in the surface density occurs at
around $N(H_2) = 6\times10^{21}$ \c2, or roughly, A$_v$ = 6, suggestive
of a threshold for star formation.  Note that the same trend is
visible even in a sample of mostly T-Tauri stars (lower right panel).

In Taurus, neither the dispersion of gas due to star formation nor the
dispersion of stars due to stellar motion is likely to have altered
the collocation of very dense gas and highly extincted young stars.
The threshold in column density for star formation is consistent
with the conclusion of \citet{mizuno1995} with the difference being
our finding a higher threshold of $6\times10^{21}$ \c2\ instead of
$3\times10^{21}$ \c2.  Given the larger number of pre--main sequence stars
available for the present work, the significance of the change in the
stellar surface density is also higher.

With our rather complete coverage of gas and stars, we can examine the
relationship of the stellar mass to the gas mass, which defines the
star formation efficiency (or SFE).  
From a very simplified point of view of the time
evolution of the star formation process, we can define the star
formation efficiency in three ways.  In the first, the SFE is defined as
the mass of all known young (pre--main sequence) stars divided by the
total gas mass.  Assuming an average mass of 0.6 solar mass for each
of the stars in our sample \citep[following][]{palla2000} 
and the total molecular mass of 2.4$\times10^4$ \Ms\
(Table \ref{mask_mass}), the star formation efficiency thus defined is
0.6 percent.

In the second, we define the SFE more strictly for the current epoch,
i.e., counting only the mass of protostars and of dense gas (that in
our mask 2 region).  The SFE thus defined in this more restricted
sense is about 0.3 percent.  For the third method, we adopt a less physically motivated
but procedurally simple approach of defining the star formation
efficiency to be the mass of all pre--main sequence stars divided by
the mass of dense gas, we obtain an SFE equal to 1.2 percent.  These
low values confirm that Taurus is a region of relatively low star
formation efficiency.

Since star formation is an ongoing process in Taurus the SFE
as defined will evolve with time.  A more meaningful quantity
is the star formation rate per unit molecular gas mass.
The star formation history of Taurus is a topic of some
controversy \citep [cf.][]{palla2000, hartmann2001, palla2002}, 
particularly the issue regarding whether the
star formation rate is presently  
accelerating or has already reached a peak and is declining.
Nevertheless, there does seem to be agreement that 
star formation has been rapid.  Star formation in Taurus 
began over 10 Myr ago, but most of the identified pre-main 
sequence stars have formed in the past 3 Myr \citep{palla2002}.  
The average star formation rate over the past 3 Myr within the region 
of Taurus included in this study has been $\simeq$ 8$\times10^{-5}$ stars yr$^{-1}$.  

Assuming as before an average mass of 0.6 solar masses, we derive
a star formation rate of 5$\times10^{-5}$ \Ms\ yr$^{-1}$.  Thus, the star formation rate
per unit molecular gas mass is approximately 2$\times10^{-9}$ \Ms\
per year per solar mass of molecular gas.  If this rate were to
continue, the gas consumption timescale would be over 400 Myr.  However, 
most of the dense gas is likely to be dispersed by
the winds from the newly forming stars long before a significant
fraction of the cloud mass is converted into stars.  It is intriguing
that the star formation rate per unit molecular gas mass in Taurus is very
similar to that found globally in the Milky Way (assuming a total 
molecular mass of 2$\times10^9$ \Ms\ and a star formation rate of 3 \Ms\ yr$^{-1}$).  

%
%
\section{MORPHOLOGY OF THE MOLECULAR GAS}
\label{morphology}

\subsection{General Structure of the Gas}

\subsection{Regions of Interest}

In this section we discuss several of the regions of particular
interest that stand out in the carbon monoxide emission from Taurus.
These are to some degree reflections of the complex structure seen on
a large scale, but highlight some of the varied structures that can
easily be identified.  The present discussion is by no means complete
but does illustrate the varied and complex structures found in
this region in which only low mass star formation is taking place.
These are grouped together by location within the cloud so that they
can be highlighted by detailed images, but this does not necessarily
reflect any physical relationship between different features.

\subsubsection{Filamentary Structure Within the Dense Gas}

A very striking feature of the molecular gas within the dense portion
of Taurus is the fact that the \th\ emission is highly structured even
in integrated intensity, as can be seen in Figure \ref{13co_tint}.  An
impressive example is shown in Figure \ref{filaments} which shows a
several approximately parallel filaments at 4\hr27\mn
+26\deg45\arcmin, having a southeast to northwest orientation.  The
filaments are $\simeq$ 20' to 25' (0.8 pc to 1.0 pc) long, with a
$\simeq$ 6:1 length to width ratio.  These filaments are readily
visible in individual velocity images \citep{narayanan2007} as well as
the \th\ integrated intensity image, but are invisible in the \tw\
data.  The peak H$_2$ column density of the filaments is
3$\times$10$^{21}$ \c2, about a factor of two greater than that of the
region between them.

Another very interesting feature visible in Figure \ref{filaments} is
the almost complete ring--like structure centered at 4\hr31\mn\
+28\deg01\arcmin.  It is fairly circular, having an angular diameter
of 18\arcmin, corresponding to 0.73 pc.  The molecular hydrogen column
density is typically 3$\times$10$^{21}$ \c2\ around the periphery of
the ring and 1.8$\times$10$^{21}$ \c2\ in the center.  This ring shows
up quite clearly in the \tw\ integrated intensity image in which $\int
T_A dv$ increases from 7.5 \kks\ in the center to $\simeq$ 11 \kks\ on
the periphery.  This features is not discernible in the \tw\ maximum
intensity image, indicating that it is showing increased line width,
although distinct kinematic structure is not evident.

%
%
\bf[!htbp]
\includegraphics[scale=0.9]{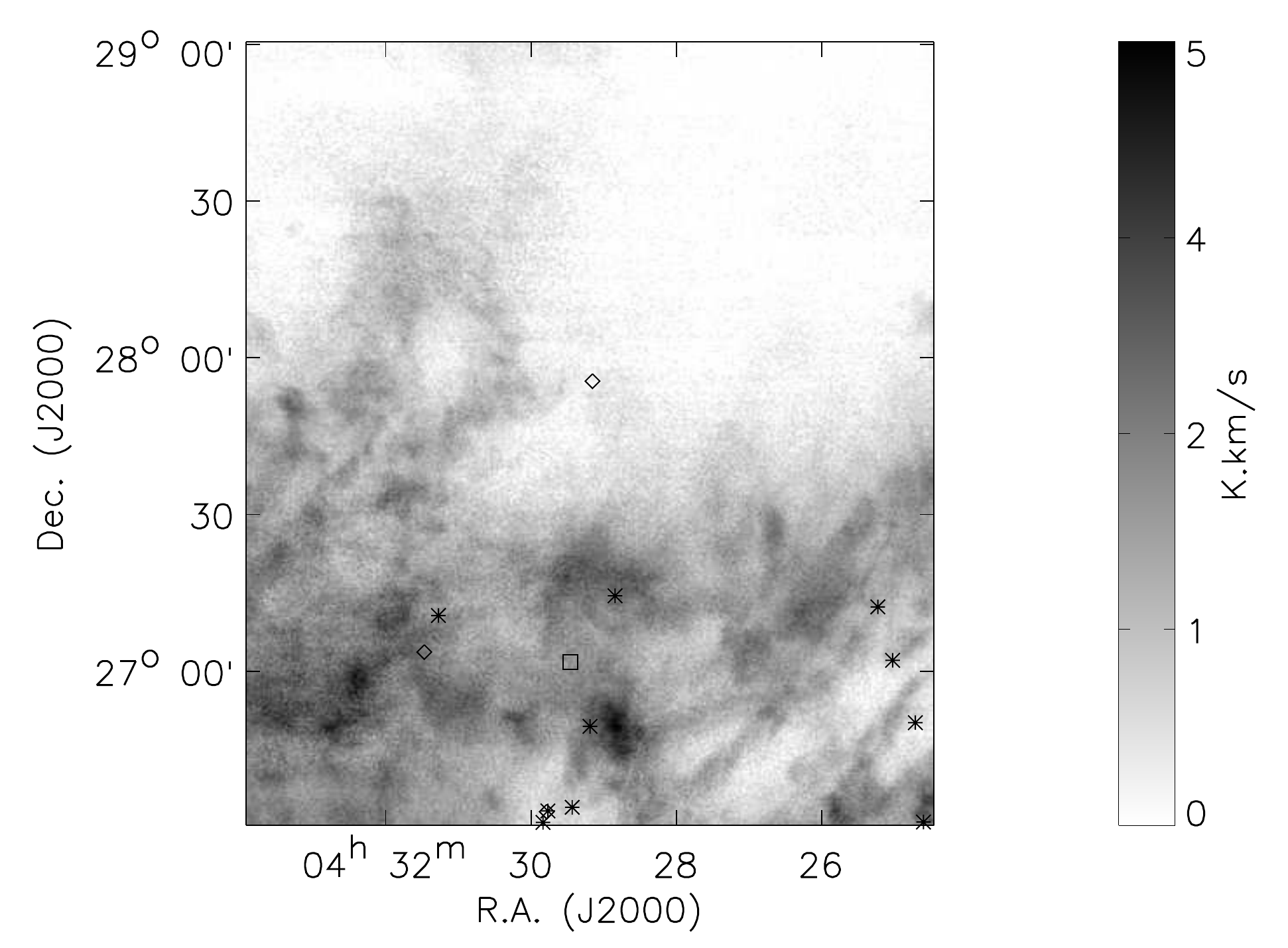}
\caption{\label{filaments} Enlarged image of \th\ integrated intensity
showing filamentary structures located near 4\hr27\mn +26\deg45\arcmin.
A nearly circular ring--like structure centered at 4\hr31\mn +28\deg\
is also visible. The young stars in this region are
indicated by the symbols defined in Fig. \ref{taurus-stars}. 
The diamonds indicate diffuse or extended young stellar sources, the 
square indicates a Class I or diffuse source, and the asterisks indicate T-Tauri stars.
}
\ef

\subsubsection{Cometary Globules and Ring in Large Cavity}

A structure that appears to be a large cavity is visible at the
eastern end of B213, just to the north of B18, visible in the \tw\
image, but more clearly in the \th\ integrated intensity (Figure
\ref{13co_tint}).  An enlarged image is shown in Figure
\ref{globules}.  The center of the cavity is approximately 4\hr29\mn
+25\deg30\arcmin.  Although the cavity is still clearly visible, it is
considerably smaller, 40\arcmin\ (1.6 pc) in \tw\ compared to
70\arcmin\ (2.9 pc) in \th.  The minimum H$_2$ column density of the
cavity is 1.4$\times$10$^{21}$ \cc\ (it is included in mask 1), but
the \th\ is detected when averaged over a reasonable number of pixels.

%
%
\bf[!htbp]
\includegraphics[scale=0.9]{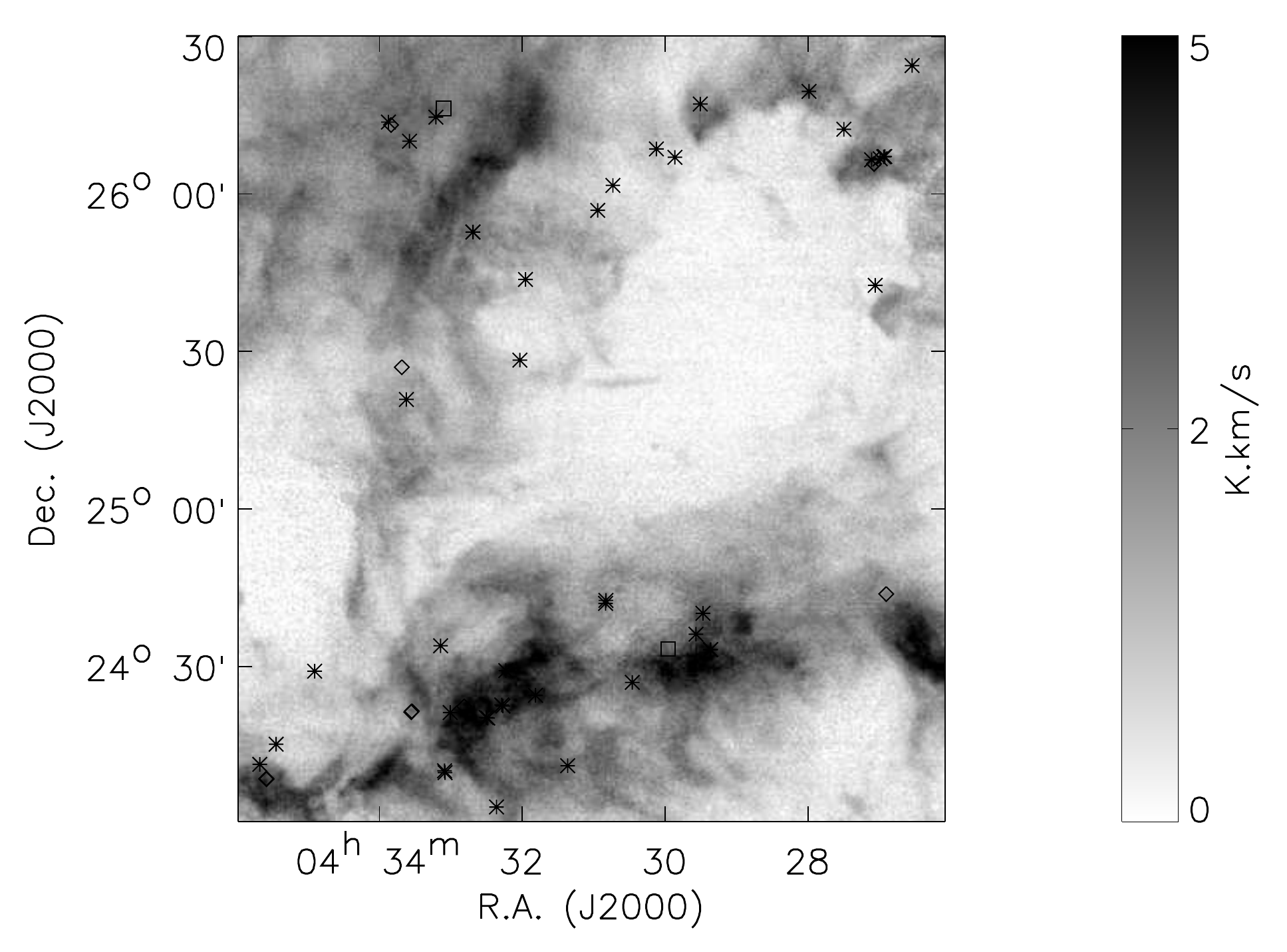}
\caption{\label{globules} This enlarged \th\ integrated intensity
image shows the cavity between B213 and B18.  At the northwest, three
cometary globules are evident, having properties presented in Table
\ref{globule_properties} and discussed in the text.  A nearly complete
ring is also seen at the eastern edge of the cavity.  
The diamonds indicate diffuse or extended young stellar sources, 
the squares indicate Class I or younger stars, and the asterisks indicate T-Tauri stars.
} 
\ef

The boundary of this cavity contains an impressive number of young stars,
which in fact nearly completely surround it.  To the north, these seem to
be distributed around the periphery of the cavity, but at its western
edge (the eastern end of B213), there are three prominent condensations,
looking remarkably like cometary globules, projecting into the cavity.
Some properties of the condensations are shown in Table
\ref{globule_properties}.  The globules are undistinguished in terms
of maximum \tw\ temperature.  The maximum column density of each of
the globules is close to 4$\times$10$^{21}$ \c2.  We have not been
able to identify any source that would be responsible for forming the
cavity, but this may be a result of its relatively great age.

\begin{deluxetable}{cc ccc}
\tablewidth{0pt}
\tablecaption{\label{globule_properties} Properties of Cometary Globules In B213--B18 Cavity}
\tablehead{	\colhead{Globule Number}&\colhead{RA(J2000)}	&\colhead{Decl(J2000)}		&\colhead{Mass}	&\colhead{Embedded Star}\\
		\colhead{}		&\colhead{}	&\colhead{}	&\colhead{\Ms}&{}\\}		
\startdata
1	&4\hr26\mn49\sc.8	&25\deg39\arcmin06\arcsec	&5.9	&DF Tau	\\
2	&4\hr27\mn06\sc.3	&26\deg06\arcmin07\arcsec	&8.3	&DG Tau	\\
{}	&{}			&{}				&{}	&FV Tau	\\
3	&4\hr29\mn25\sc.5	&26\deg14\arcmin42\arcsec	&4.1	&FW Tau	\\
\enddata
\end{deluxetable}

As indicated in Table \ref{globule_properties} (see also Figure
\ref{taurus-stars}), each of the globules contains a T Tauri star,
with Globule 2 containing two stars.  DF Tau is located slightly
inwards (toward the cavity center) relative to Globule 1, while the
stars in Globules 2 and 3 are located 3\arcmin\ away from the cavity
center compared to the tip of the globule.  There does not appear to
be any readily discernible kinematic signature giving clues to the
origin of the globules, or revealing an effect of the star formation.
For example, although the star DG Tau B in Globule 2 has an optical jet which is 
presumed to be driving the observed red--shifted molecular outflow \citep{mitchell1997},
we do not see an effect on the quiescent gas distribution.

The stars in question range from 0.2 \Ls\ to 2.2 \Ls, and have ages
between 0.6 Myr (DG Tau) to 1.2 Myr (FW Tau).  Stars of this age may
well have moved a significant distance since their formation, so that
it is not surprising that if they were formed in these globules by
e.g. radiative implosion \citep{bertoldi1990}, they may now appear
displaced from their formation sites.

\subsubsection{Irregular Filament or Boundary in L1536}

%
%
\bf[!htbp]
\includegraphics[scale=0.9]{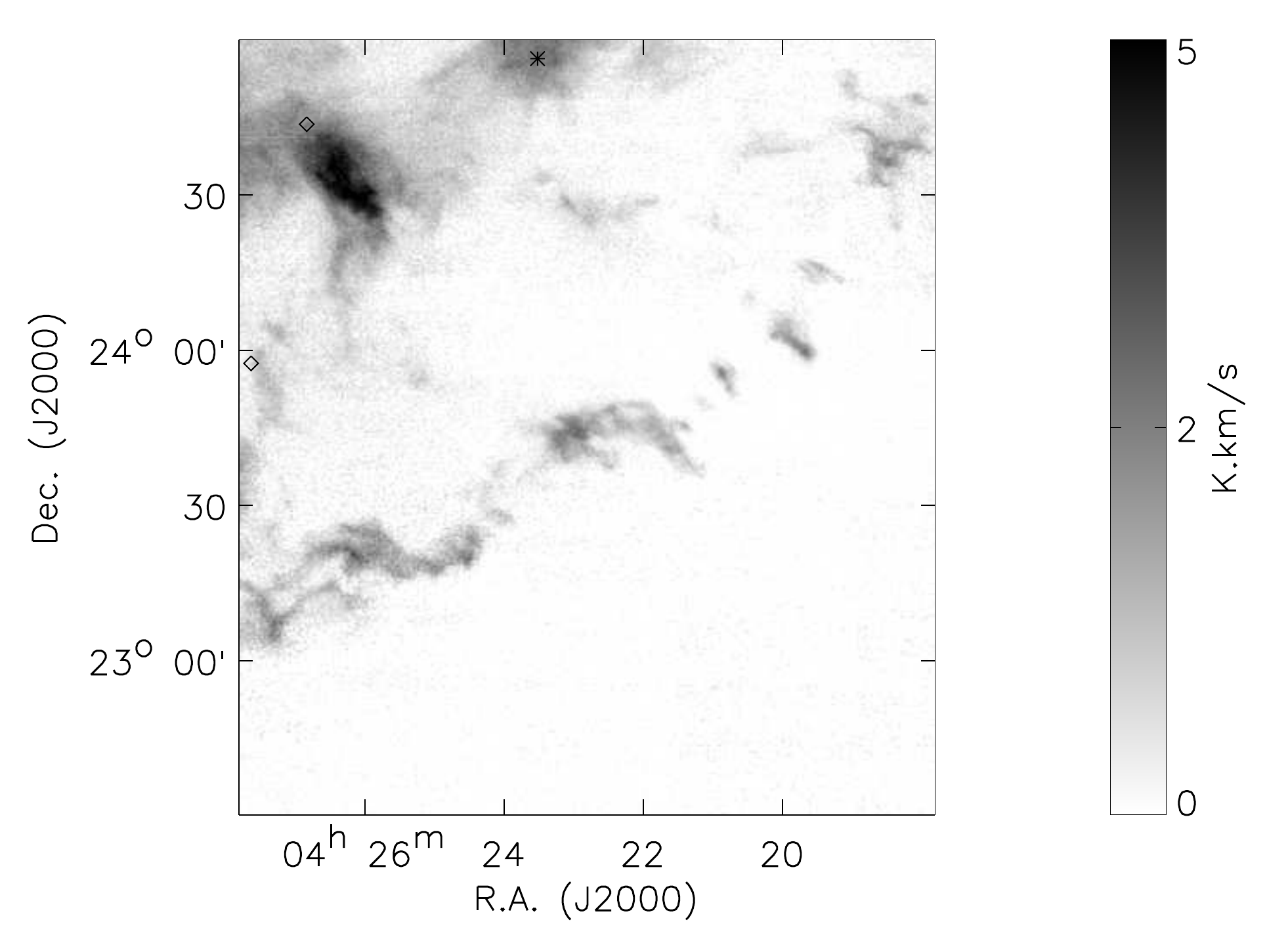}
\caption{\label{twisted_filament} Filament running to northwest from
L1536 seen in \th\ integrated emission.  The condensations have a
maximum \h2\ column density of 5$\times$10$^{21}$ \c2.
The diamonds indicate diffuse or extended young stellar sources 
and the asterisk indicates a T-Tauri star.
} 
\ef

A very long filament having one end in the south--central portion of
L1536 and extending to the northwest is visible in the \th\ emission,
shown in an enlarged view in Figure \ref{twisted_filament}.  The
filament center is at 4\hr23\mn\ +23\deg45\arcmin, and its length is
2\deg, corresponding to 4.9 pc.  The morphology of the filament is
suggestive of its being a boundary between regions of lower
(to the south) and higher (to the north) column density. 
The form of the filament is somewhat suggestive of a helix, but it could simply have an
irregular shape.  The \h2\ column density along the filament is
typically 3$\times$10$^{21}$ \c2, but reaches 5$\times$10$^{21}$ \c2
in the regions of strongest emission.  The region surrounding the
filament has a \h2\ column density of 1.3 to 1.5 $\times$10$^{21}$
\c2, only slightly greater than our minimum value defined by mask 0 of
1.1$\times$10$^{21}$ \c2.  This filament, is roughly parallel to the
structure formed by B18 and L1506, to the filamentary part of B213,
and also to the less well--defined but still quite flattened structure
formed by Heiles' Cloud 2 and L1521.  This thin filament is the most
southerly and furthest from the Galactic plane of all of these
structures.  The position angle of all four of these
filamentary/elongated clouds is approximately 45\deg\ relative to the
plane of the Milky Way.

\subsubsection{Molecular Ring and Planar Boundary}

Figure \ref{planar_boundary} includes several different structures.  The
first is the ``molecular ring'', studied in detail by
\cite{schloerb1984}.  This ring, 30\arcmin\ (1.2 pc) in diameter,
centered at 4\hr40\mn30\sc\ +25\deg45\arcmin, contains at least 6
dense condensations visible in the \th\ integrated intensity image.
The best--studied of these is the chemically very interesting TMC-1
ridge, observed in detail by \cite{pratap1997} and many others.  The
ridge (the NH$_3$ peak is at 4\hr41\mn21\sc +25\deg48\arcmin) is not
very prominent in the \th\ integrated intensity image, which is
presumably a result of the significant optical depth in the ring
material, which may not be corrected for entirely by the simple
process (described in \S \ref{col_dens_calc}) employed here.  The
peak \h2\ column density we derive is 7$\times$10$^{21}$ \c2\, which
is somewhat less than half of that which would be derived from the
\ce\ observations of \cite{pratap1997}.  Given the difficulties
expected in deriving the column density in regions of optically thick
emission in which significant temperature gradients may be present,
this difference is not unreasonable.  The ridge is more visible in our
\tw\ map than in that of \cite{schloerb1984} due to the better
sampling in the present work.

%
%
\bf[!htbp] 
\includegraphics[scale=0.9]{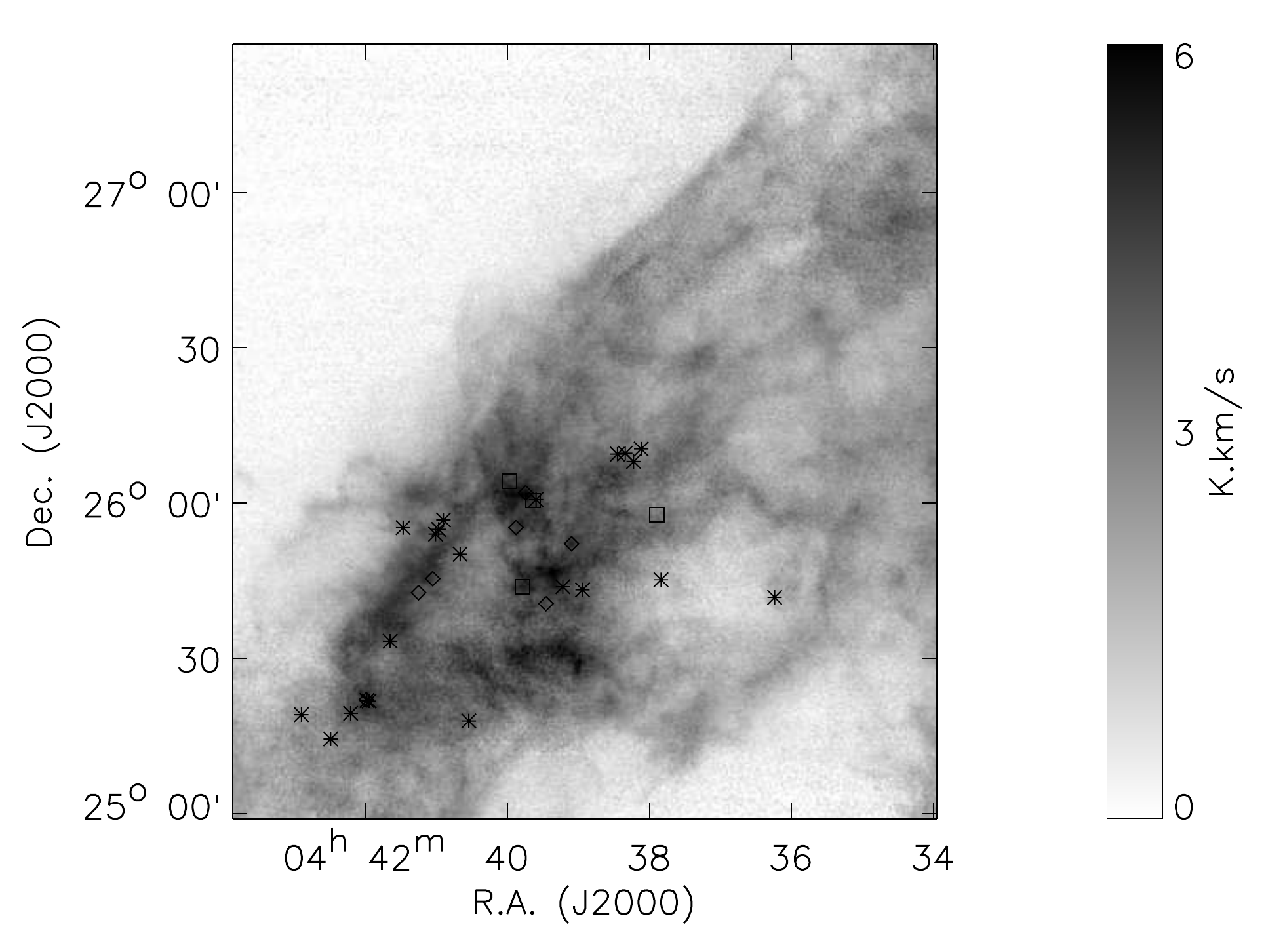}
\caption{\label{planar_boundary}\th\ integrated intensity image
showing the molecular ring in Heiles' Cloud 2 centered at
4\hr40\mn30\sc\ +25\deg45\arcmin.  We also see the very well--defined
planar boundary centered at 4\hr38\mn30\sc\ +26\deg50\arcmin.  The
\tw\ boundary in this region is not at all sharp, having a hair-like
appearance as seen most clearly in Figure \ref{CO+B}.  
The diamonds indicate diffuse or extended young stellar sources, 
the squares indicate Class I or younger stars, and the asterisks indicate T-Tauri stars.
} \ef

The second noticeable feature in Figure \ref{planar_boundary} is the
very straight boundary of the molecular emission seen in \th\,
centered at 4\hr38\mn30\sc\ +26\deg50\arcmin\ and extending for over 1
degree (2.4 pc).  The questions of the formation of this interface and
how it is maintained are intriguing.  In this region, the \tw\
emission extends significantly beyond that of the \th\ away from the
high column density portion of the cloud, typically by 0.5 pc.  As can
be seen in Figures \ref{12co_tmax} and \ref{CO+B}, the \tw\ emission
is highly structured, particularly perpendicular to the interface
direction.  This behavior is not restricted to this portion of the
cloud boundary, but in fact is a general characteristic of the \tw\
emission in the mask 1 region surrounding the high column density
portion of the cloud (mask 2) where \th\ is detected in individual
spectra.

Finally, we note the intriguing feature to the west of the better--known ring
discussed above.  With a center at 4\hr37\mn +26\deg45\arcmin, this is 
again a slightly non circular ring having a diameter of 30\arcmin\ (1.2 pc).
Given the complexity of the structure observed in our study of the molecular
gas in Taurus, this could certainly be a superposition of filaments rather than a ring.

\subsubsection{L1495 and B213}

The L1495 region contains the greatest concentration of young stars
within the region of the Taurus molecular cloud that we have mapped.  
Figure \ref{L1495} shows the eastern part of L1495; the western part (seen in Figure
\ref{13co_tint}) is more diffuse.  The enlarged image also shows the
very narrow B213 filament which extends to the southeast from L1495.
The \th\ emission and the \h2\ column density we derive from it, are
relatively continuous over the high column density portion of L1495
and the B213 filament.  In \ce\ \citep{onishi1996} individual dense
cores are better resolved, and in HCO$^+$ \citep{onishi2002} they
stand out yet more clearly.

The central part of of L1495 contains over 20 young stars in Palla's
compilation \citep{palla2008}, and has a maximum \h2\ column density
of 10$^{22}$ \c2, which is the highest we see in our map.  The mass of
the L1495 region is (Table \ref{roi_masses}) 2.6$\times$10$^3$ \Ms,
but a significant fraction of this is in the spatially extended, lower density
material.

The B213 filament is approximately 75\arcmin\ or 3 pc in length, and
only 4.5\arcmin\ or 0.2 pc thick.  One of the curious features about
this structure is that while there are dense cores seen along its
entire length \citep{onishi1996,onishi2002}, young stars have
apparently not yet formed in the northwestern 30\arcmin\ (1.2 pc) long
portion closest to L1495.  The magnetic field orientation at the
boundaries of this filament is strikingly oriented perpendicular to
its long axis, as seen dramatically in Figure \ref{CO+B}, and
discussed in \S \ref{magnetic}.

%
\bf[!htbp]
\includegraphics[scale=0.9]{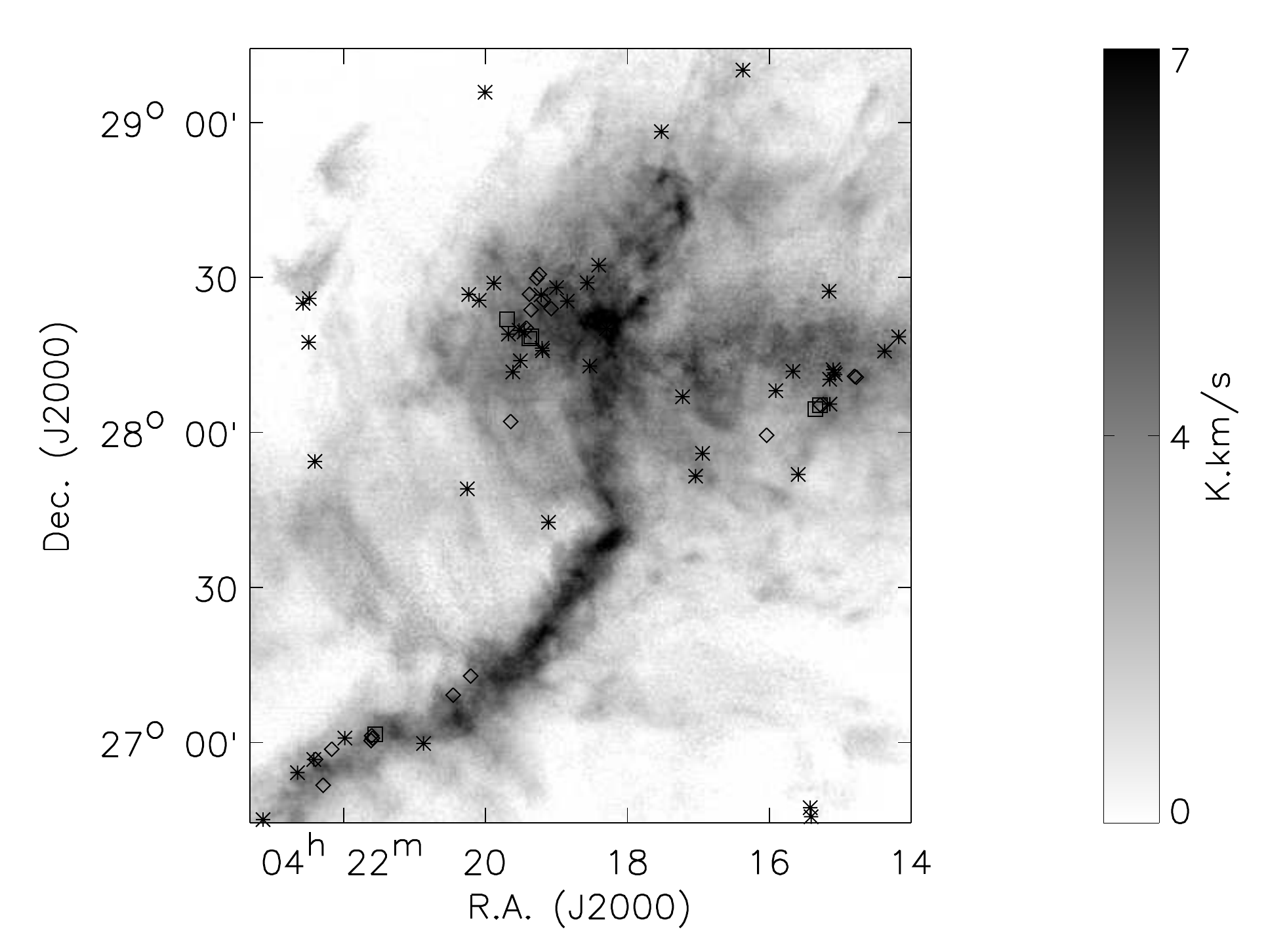}
\caption{\label{L1495} Enlarged \th\ integrated intensity image of the
western portion of L1495 (4\hr 17\mn 30\sc\ +28\deg 20\arcmin) and the
filamentary region B213 (running to the southeast from 4\hr 17\mn
30\sc\ +27\deg 40\arcmin.  
The diamonds indicate diffuse or extended young stellar sources, 
the squares indicate Class I or younger stars, and the asterisks indicate T-Tauri stars.
} 
\ef

\subsubsection{Striations in \tw\ Emission}
\label{striations}

One of the surprising features in the map of \tw\ is the prominent
striations (or threads, or strands) seen in the lower level emission
seen away from the main molecular condensations.  These can be
recognized in Figure \ref{12co_tmax}, but this effect is more visible in the
enlarged image shown in Figure \ref{12co_striations}.  Another region
in which this is very prominent is located at 4\hr15\mn\
+24\deg30\arcmin.  These are similar to structures seen within some
infrared cirrus clouds.  The striations are visible in images of
maximum antenna temperature and also integrated antenna temperature.
The characteristic values are $T_A$ = 3 K on the striations and 2 K
between them, while $\int T_A dv$ drops from $\simeq$ 2.8 \kks\ on the
striations to between 1 and 1.5 \kks\ between them.  Given that the
density in these regions is low, the \tw\ emission is almost certainly
subthermally excited so that it is difficult to determine the kinetic
temperature.  Based on the procedure described in \S 
\ref{col_dens_calc}, which assumes $T_{kin}$ equal to 15 K, the \h2\
column density of the striated features is 2$\times$10$^{21}$ \c2,
approximately double that of the background emission.  A striking
feature of the striations is their alignment parallel to the direction
of the magnetic field measured by optical starlight polarization, as shown
in Figure \ref{CO+B} and discussed in \S \ref{magnetic}.
%
%
\bf[!htbp]
\includegraphics[scale=0.9]{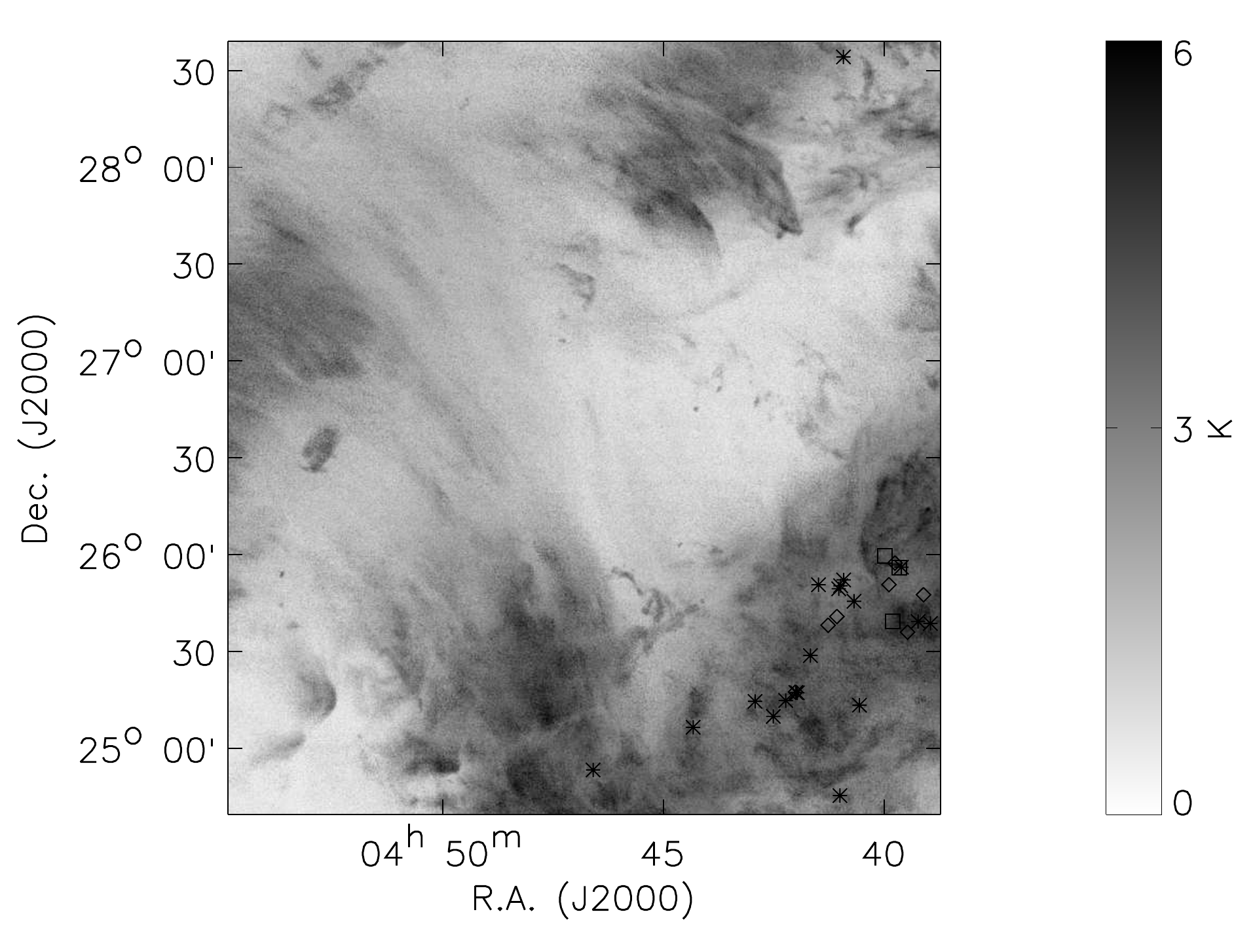}
\caption{\label{12co_striations} One of the regions showing prominent
striations in the \tw\ emission, displayed here in an image of the
maximum \tw\ antenna temperature.  
The diamonds indicate diffuse or extended young stellar sources, the 
squares indicate Class I or younger stars, and the asterisks indicate T-Tauri stars.
} 
\ef
%
%

\section{SUMMARY AND CONCLUSIONS}
\label{summary}

We have carried out a large--scale survey of the molecular gas in 
Taurus by mapping a 100 square degree region with the 13.7 m
Five College Radio Astronomy Observatory millimeter telescope.  The J = 1
$\rightarrow$ 0 transition of \tw\ and of \th\ were observed
simultaneously using the 32 pixel Sequoia focal plane array receiver.
The observing and data reduction techniques are discussed by
\cite{narayanan2007}.  In this overview, we have discussed some of the
highlights of the data that we have obtained, deferring detailed
analyses to future papers.

The combination of an unbiased, high sensitivity survey with coverage
of a relatively large area allows us to study the structure and
properties of the molecular gas in new ways.  With approximately 3
million independent spatial pixels, we have a linear dynamic range
which is unequaled in previous studies of the Taurus region.  While
our angular resolution is inferior to that obtained with larger/higher
frequency telescopes or interferometers, the strength of the present
work is to show the relationship between structures on scales ranging
from $\simeq$ 1\arcmin\ or 0.04 pc to 10 degrees (approximately 25
pc).  Our observations are sensitive to a range of column densities
equivalent to a range in visual extinction between 1 and 10
magnitudes.

\textbf{Cloud Morphology} One of our key conclusions is that the morphology of this region is very complex.  In contrast to earlier large--scale surveys carried out
with low angular resolution in which clouds appeared largely
smooth--edged and having little structure, we find an astoundingly
rich range of structures including filaments, ridges, blobs, and
holes.  The internal structure is more striking in \th\ than in
\tw\, which is not surprising given the large optical depth of the
former isotopologue.  The filaments have lengths up to 3 pc, and axial
to transverse dimension ratios as large as 15:1.  Holes in the
molecular emission appear on a large range of scales extending from
0.1 pc to 3 pc.
 
The edges of the dense molecular regions are generally very irregular,
with structures on the order of 0.1 pc in size visible especially in
\tw\ which traces cloud boundaries which are more extended than seen
in the \th.  This ``hair--like'' edge structure is found to be common
in \tw\, while the \th\ cloud boundaries are relatively sharper but
still quite irregular.  There is one notable exception in which we
find a sharp, straight boundary in \th\ almost 2.5 pc in length.

\textbf{Cloud Mass and Mass Distribution} Having both the \tw\ and \th\ detected in regions of relatively large column density (mask 2, comprising about 1/3 of the map pixels), we
have used the standard method to derive the kinetic temperature and
molecular column density, including a correction for saturation of the
\th\ which becomes significant for the regions of greatest column
density.  To analyze portions of the image (mask 1 comprising about
1/3 of total area of the cloud mapped) in which we detect \tw\ but not
\th\ in individual pixels we use a different approach.  With $\simeq$
1 million such pixels available, we have binned them by \tw\
excitation temperature $T_{ex}$.  When spectra within a bin are averaged,
the \th\ as well as the \tw\ is readily detectable, and we obtain the
\h2\ density and the CO column density.  We thus have a relationship which
gives us n(H$_2$) and N(CO) as a function of $T_{ex}$(\tw).  Since the excitation
temperature is available for each pixel, we can derive the CO column
density for each line of sight.  Averaging together all the pixels in mask 0
(in any one of which neither \tw\ nor \th\ was detectable), we detect both
isotopologues, and use the two spectra to derive the average density
and column density for mask 0, the final third of the map.  This procedure
allows us to determine the CO column density throughout the region
mapped, including even regions of relatively low column density.

To convert $N(CO)$ to total column density, we have used the results of
\cite{vandishoeck1988} which are appropriate for Taurus.  The
essential point is that the fractional abundance of carbon
monoxide drops as the total \h2\ column density is reduced, as a
result of reduced dust shielding and self--shielding.  Inverting this
argument, the column density of \h2\ corresponding to a low column
density of carbon monoxide is larger than would be obtained assuming a
constant fractional abundance for CO.  The result is that the total
mass for the region of Taurus mapped is close to 2.4$\times$10$^4$
\Ms, compared to less than 1$\times$10$^4$ \Ms\ that would be found using a
standard, uniform fractional abundance.  We find that half the mass of
the cloud is contained in regions having column density below
2.1$\times$10$^{21}$ \c2.  This result reduces the fraction of mass
found in dense cores by a factor greater than 2, and also confirms the
presence of significant external pressure in the regions external to
the dense regions.  
The total mass for the region we have mapped thus obtained agrees well with that predicted from the CO luminosity,
5.55$\times$10$^3$ K km s$^{-1}$ pc$^2$, and a standard conversion M(\Ms) = 4.1 L$_{CO}$ (K km s$^{-1}$ pc$^2$).  
It seems likely that our conclusion that a significant component of diffuse
molecular gas accompanies the more widely studied high density regions
is not restricted to Taurus. It reinforces the importance of 
observations which can study this diffuse molecular material, which is not readily detected in individual spectra with the sensitivity typically available in large--scale
molecular cloud surveys.

\textbf{Cloud Structure and Star Formation} The structural complexity over a wide range of scale sizes hints at the richness of the physical processes which underly the formation and
evolution of molecular cloud complexes such as Taurus.  The present
data set, both in terms of morphology and mass distribution,
constitutes a potentially valuable resource for comparison with
outputs from simulations of cloud formation.  The large scale
kinematic structure that we see confirms that identified in earlier
studies.  Along with the complexity of the line profiles observed
along many lines of sight, this poses a real challenge for any
detailed theoretical model of this region.

We see a varied relationship between the magnetic field as measured by
polarization of background stars, and the distribution of the gas.  In
the more diffuse regions traced by \tw\, we see large--scale alignment
between the field direction and striated structure in the gas.
Although we have not been able to measure any kinematic signature, the
appearance is strongly suggestive of flows along the field lines.  In
several of the very elongated filaments seen in the denser gas traced
by \th, the magnetic field is oriented perpendicular, or nearly
perpendicular, to the major axes of the filaments.  Combined with the
hair--like appearance of the boundaries of these filaments seen in
\th\, but more prominent in \tw, this again suggests that motions of
material along the field lines have been responsible for building up
the regions of higher density within the overall molecular cloud.

The surface density of very young and moderately young stars shows a 
rapid increase at a H$_2$ column density of 6$\times$10$^{21}$ \c2,
confirming the existence of a threshold for star formation.
We have used new compilations of young stars in the Taurus region to
calculate the star formation efficiency (SFE).  Our large value for
the gas mass, especially in regions of lower column density, results
in the SFE, taken to be the mass of all young stars in the region
divided by the total molecular mass, to be 0.6 percent.  Taking the SFE
for most recent star formation by comparing the mass of only the
embedded protostars with that of the dense gas, gives an SFE equal to
0.3 percent.  If we consider all of the young stars (whether embedded
protostars or T-Tauri stars) in the region of high column density, we
obtain a SFE equal to 1.2 percent.
The average star formation rate over the past 3 Myr within the region 
of Taurus included in this study has been $\simeq$ 8$\times10^{-5}$ stars yr$^{-1}$,
corresponding to a mass going into new stars of 5$\times10^{-5}$ \Ms\ yr$^{-1}$.

This work was supported in part by the National Science Foundation
through grant AST-0407019 to Cornell University, and by the Jet
Propulsion Laboratory, California Institute of Technology.  
The Five College Radio Astronomy Observatory is operated with support 
from the National Science Foundation through NSF grant AST 05 40852 and with
permission of the Metropolitan District Commission.  We thank
Yvonne Tang for contributions to data taking and analysis of dense
condensations in Taurus, and Marko Krco for assistance with
observations.  We thank Pierre Hily--Blant for the suggestion to compare
the magnetic field and integrated intensity maps in Taurus, and for 
many useful conversations about this and other topics.  
We are indebted to Francesco Palla and Scott Kenyon for
providing compilations of young stars in the Taurus region and their
properties.  We thank Ted Bergin for carrying out time--dependent calculations of
the CO abundance in diffuse regions.  We thank the anonymous reviewer for very carefully reading the lengthy manuscript, noting some problems, and making some suggestions for 
further work which has improved this study.  This research has made use of 
NASA's Astrophysics Data System.



\end{document}